\documentclass{JHEP3}
\pdfoutput=1
\usepackage{graphics}
\usepackage{graphicx}
\usepackage{amsmath}
\usepackage{cite}
\def\be{\begin{equation}}
\def\ee{\end{equation}}
\def\bea{\begin{eqnarray}}
\def\eea{\end{eqnarray}}
\def\eps{\epsilon}
\def\nnb{\nonumber}
\def\eps{\epsilon}
\def\dps{\displaystyle}

\def\cN{  {\cal N}  }
\def\cO{  {\cal O}  }

\preprint{
SI-HEP-2013-03}
\title{The four-loop cusp anomalous dimension in $\cN=4$ super Yang-Mills
and analytic integration techniques for Wilson line integrals}

\author{Johannes M.\ Henn$^{a}$, Tobias Huber$^{b}$\\
$^a$ Institute for Advanced Study, Princeton, NJ 08540, USA\\
$^b$ Theoretische Physik 1, Naturwissenschaftlich-Technische Fakult\"at,
\\ Universit\"at Siegen, Walter-Flex-Strasse 3, D-57068 Siegen, Germany\\

\email{jmhenn@ias.edu, \\ huber@tp1.physik.uni-siegen.de}}

\abstract{
Correlation functions of Wilson lines are relevant for describing the infrared
structure of scattering amplitudes.
We develop a new method for evaluating a wide class of such Wilson line integrals,
and apply it to the calculation of the velocity-dependent cusp anomalous dimension in 
maximally supersymmetric Yang-Mills theory.
We compute the four-loop non-planar correction in a recently introduced scaling limit.
Moreover, we derive the full planar four-loop result by means of an ansatz
which is based on the structure of known analytic results.
We determine the coefficients in this ansatz by making use of a relationship
to massive scattering amplitudes.
As a corollary, our analytical result confirms the four-loop value of the light-like cusp
anomalous dimension.
Finally, we use the available perturbative data, as well as insight from AdS/CFT, in order to extrapolate
the leading order values at strong coupling. The latter agree within two per cent with the corresponding
string theory result, over a wide range of parameters.
}


\begin{document}

%%%%%%%%%%%%%%%%%%%%%%%%%%%%%%%%%%%%%%%%%%%%%%%%%%%%%%%%%%%%%%%%%%%%%%%%%%%%%%%%%%%%%%%%%%%%%%%%%%%%%%%%%%%%%%%%%%%%%%%%%%%%%%%%%%%%%%%%%%%%%%%%%%%%%
%%%%%%%%%%%%%%%%%%%%%%%%%%%%%%%%%%%%%%%%%%%%%%%%%%%%%%%%%%%%%%%%%%%%%%%%%%%%%%%%%%%%%%%%%%%%%%%%%%%%%%%%%%%%%%%%%%%%%%%%%%%%%%%%%%%%%%%%%%%%%%%%%%%%%

\section{Introduction}
\label{sec-intro}

Wilson loops are very fundamental and important quantities in gauge theories. 
In this paper, our main focus will be on their relevance to the description of the
infrared (IR) behavior of scattering amplitudes. We will consider the general
massive case, from which the massless one can be obtained as a limit.

The appearance of Wilson loops in this problem is easy to understand. The infrared divergences
in scattering amplitudes originate from soft regions of loop integration, for which one
can employ the eikonal approximation. In this way, one finds that infrared divergences
of a scattering process are given by a correlation function of Wilson lines,
where the lines in position space point along the momenta of the scattered particles.
However, in taking the eikonal limit, additional ultraviolet (UV) divergences are introduced.
They are equivalent, up to a sign, to the original IR divergences. 
This allows one to regard the former as the UV anomalous dimension of Wilson line operators, 
whose renormalization properties are well understood \cite{Polyakov:1980ca,Brandt:1981kf,Korchemsky:1985xj}.
Note that the latter depends on the color representation of the external particles and
is in general a matrix in color space. It is known analytically to two loops \cite{Ferroglia:2009ii}.
The analysis of the general structure of this soft anomalous dimension matrix is of great interest,
with recent studies involving the massless~\cite{Gardi:2009qi,Dixon:2009gx,Becher:2009qa,Dixon:2009ur,Ahrens:2012qz}
and massive~\cite{Becher:2009kw} case.

In the planar limit, the matrix factorizes into Wilson lines consisting of two segments.
The cusp anomalous dimension associated to two Wilson lines is known in QCD to two
loops \cite{Korchemsky:1987wg}, and in $\mathcal{N}=4$ supersymmetric Yang-Mills (SYM) to three loops \cite{Correa:2012nk}.
In this paper, we extend the calculation in planar $\mathcal{N}=4$ SYM to four loops,
and, in addition, compute the non-planar four-loop value in a  special scaling limit.

The aim of this paper is to develop methods for the computation of such Wilson line correlators,
planar and non-planar, and to deepen the understanding of the functions involved.
This is closely related to ideas being discussed for understanding the loop corrections to
scattering amplitudes. 
The functions that are typically encountered can be described by certain classes
of iterated integrals. A key problem is to identify which specific class of functions is required
to describe a given scattering process.
It was found that integrals for scattering amplitudes or Wilson loops can be put into a `d-log' form~\cite{CHTrento,ArkaniHamed:2012nw,Lipstein:2012vs}, where one can pull out an overall normalization
factor, and the remaining integrand is a differential form. Moreover, such a representation suggests
the existence of simple differential equations for the integrals. 
The latter also help to make the transcendentality properties of the integrals manifest.
Recently, evidence was presented that integrals having such simple properties are
not limited to supersymmetric theories, but can be present in generic $D=4-2 \eps$ dimensional
integrals \cite{Henn:2013pwa}.
The Wilson line integrals considered in this paper can be considered as a special, simplifying
limit of the more general scattering amplitude integrals.
We will derive `d-log' representations for a wide class of Wilson line integrals, relevant
to the physical problems discussed above, and show how to compute them using differential 
equations.

The $\mathcal{N}=4$ supersymmetric Yang-Mills (SYM) theory is a good testing ground for
exploring such Wilson loops. For the specific Wilson loops studied in this paper, 
results can be obtained from various methods such as supersymmetric localization techniques or, in the planar case, integrability~\cite{Correa:2012hh,Drukker:2012de}.
The AdS/CFT conjecture also allows to compute Wilson loops at strong coupling.

\FIGURE[t]{
\includegraphics[width=0.65\textwidth]{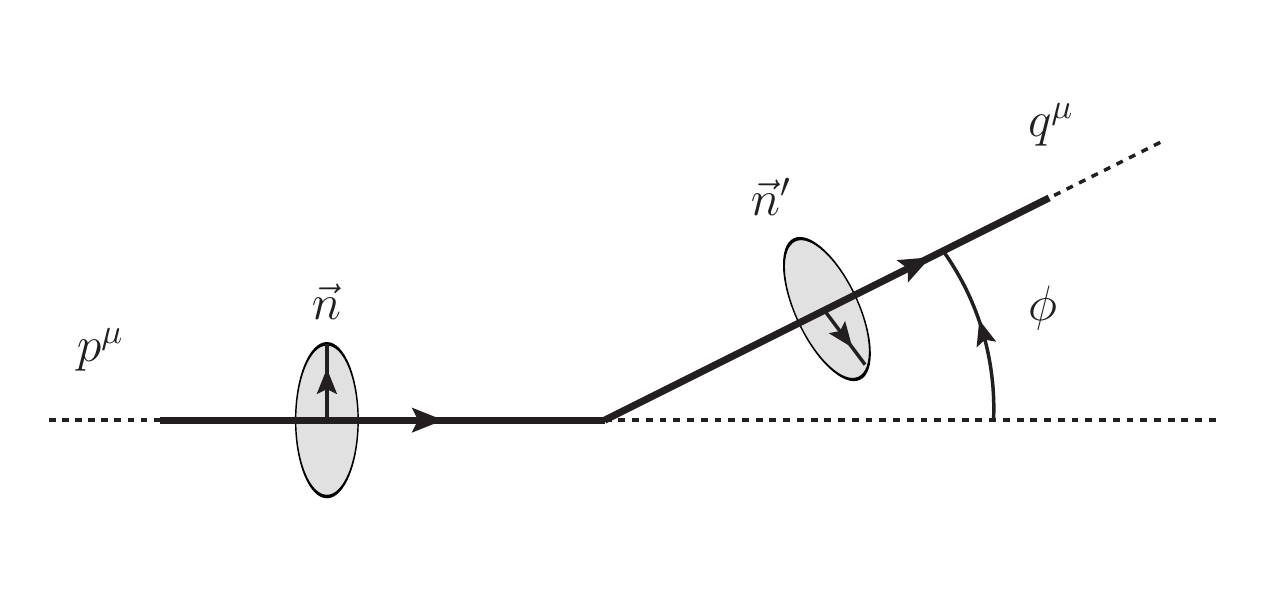}
\caption{A Wilson line that makes a turn by an angle $\phi$ in Euclidean space.
The two segments go along $p^{\mu}$ and $q^{\mu}$, respectively.
The vectors $\vec{n}$ and $\vec{n}'$ are internal vectors that determine the
coupling to the six scalars $\vec{\Phi}$, see eq. (\ref{susy-loop}).}\label{fig:WL}}

In $\mathcal{N}=4$ SYM, it is natural to define the locally supersymmetric
Wilson loop operator \cite{Maldacena:1998im,Rey:1998ik}
\begin{eqnarray}\label{susy-loop}
W \sim {\rm Tr}\left[P \exp \left({ i \oint  A^{\mu} \dot{x}_{\mu} + \oint |dx| \vec{n} \cdot \vec{\Phi}} \right) \right]\,,
\end{eqnarray}
where $\vec{n}$ is a vector on $S^{5}$. It parametrizes the 
coupling of the Wilson loop to the six scalars $\vec{\Phi}$.
We consider as the integration contour a cusp formed by two segments
along directions (momenta) $p^{\mu}$ and $q^{\mu}$,
and allow the two segments to couple to the scalars through $\vec{n}$ and $\vec{n}'$, see Fig.~\ref{fig:WL}.
Then, the vacuum expectation value $\langle W \rangle$ of the Wilson loop will depend on the angles
\begin{eqnarray}
\cos \phi  = \frac{ p \cdot q}{\sqrt{p^2 q^2}} \,,\qquad  \cos \theta = \vec{n} \cdot \vec{n}' \,,
\end{eqnarray}
as well as on the `t Hooft coupling $\lambda=g^2 N$, and the number of colors $N$.

If $\Lambda_{\rm UV}$ and $\Lambda_{\rm IR}$ are short and long distance cutoffs, respectively,
then the divergent part of the vacuum expectation value of the Wilson loop takes the form \cite{Polyakov:1980ca,Brandt:1981kf}
\begin{eqnarray}\label{eq:defcuspAD}
\left\langle W \right\rangle \sim \exp{ \left[ - \log \frac{\Lambda_{\rm UV }}{\Lambda_{\rm IR}} \, \Gamma_{\rm cusp} + \ldots \right] } \,.
\end{eqnarray}
This defines the cusp anomalous dimension $\Gamma_{\rm cusp}(\phi, \theta, \lambda, N)$.

Note that the dependence of $\Gamma_{\rm cusp}$ on $\theta$ is simple. 
It can only occur through Wick contractions of scalars, and because of SO(6) invariance
it appears only through $\vec{n} \cdot \vec{n}' = \cos \theta$. 
Therefore, at $L$ loops, $\Gamma_{\rm cusp}$ is a polynomial in $\cos \theta$,
of maximal degree $L$. 
Having made this observation, we find it convenient
to introduce the variable 
\begin{align}
\xi = \frac{\cos \theta - \cos \phi}{i \sin \phi} \,,
\end{align}
where the denominator was chosen for future convenience.
When the geometric angle $\phi$ and internal angle $\theta$ satisfy $\phi = \pm \theta$, which corresponds to $\xi \to 0$,
the anomalous dimension vanishes. 
Thus we expect the following structure in perturbation theory,
\begin{align}\label{gamma-general}
\Gamma_{\rm cusp}(\phi,\theta,\lambda,N) = \sum_{L=1}^{\infty}  \left( \frac{\lambda}{8 \pi^2}\right)^L \sum_{r=1}^{L}\xi^r \, \Gamma^{(L;r)}(\phi,1/N^2)  \,,
\end{align}
where $\lambda = g^2 N$ is the `t Hooft coupling, and $g$ the Yang-Mills coupling.
The sum over $\xi^r$ starts from $r=1$, since $\xi=0$ corresponds to a supersymmetric configuration, for which $\Gamma_{\rm cusp}$ vanishes.

Note that $\Gamma_{\rm cusp}$ has non-planar corrections starting from four loops.
We will discuss the full structure of the color dependence to four loops in sections \ref{sec-color} and  \ref{sec-ladders}.

The $r=1$ term is known to all loop orders, including the non-planar corrections~\cite{Correa:2012at}.
In this paper, we will compute the full planar result at four loops, for $\theta=0$, as well as the non-planar contribution to $\Gamma^{(4;4)}$, which
is the leading term in the scaling limit $\xi \to \infty$. This is done by analytically continuing $\theta$ and keeping $\phi$ as a free parameter. 
This scaling limit was introduced in ref. \cite{Correa:2012at} and it was shown that it allows to describe the planar ladder
diagrams in a simple way. This was further developed in \cite{Bykov:2012sc,Henn:2012qz}.

This paper is organized as follows. In section~\ref{sec-color} we review the color dependence of $\Gamma_{\rm cusp}$ to four loops.
In section~\ref{sec-kinematics} we explain the kinematics and give an overview of the
functions that will appear in $\Gamma_{\rm cusp}$. 
In section~\ref{sec-ladders} we discuss the structure of a class of Wilson line integrals and
propose a systematic way of evaluating them.
We then apply this to the non-planar correction to $\Gamma_{\rm cusp}$ in the scaling limit. 
In section~\ref{sec-amplitude} we give details on how we compute the planar
$\Gamma_{\rm cusp}$ from a massive scattering amplitude, 
and give the result for the four-loop cusp
anomalous dimension in the planar limit. 
In section~\ref{sec-strong}
we compare our results to those at strong coupling. 
We conclude in section~\ref{sec-discussion}.
There are two appendices. 
In appendix~\ref{app:integrals} we collect the contributing diagrams from the scattering amplitude and discuss their structure, while
appendix~\ref{app:analyticcont} contains the analytic continuation of $\Gamma_{\rm cusp}$ to values beyond threshold.

%%%%%%%%%%%%%%%%%%%%%%%%%%%%%%%%%%%%%%%%%%%%%%%%%%%%%%%%%%%%%%%%%%%%%%%%%%%%%%%%%%%%%%%%%%%%%%%%%%%%%%%%%%%%%%%%%%%%%%%%%%%%%%%%%%%%%%%%%%%%%%%%%%%%%
%%%%%%%%%%%%%%%%%%%%%%%%%%%%%%%%%%%%%%%%%%%%%%%%%%%%%%%%%%%%%%%%%%%%%%%%%%%%%%%%%%%%%%%%%%%%%%%%%%%%%%%%%%%%%%%%%%%%%%%%%%%%%%%%%%%%%%%%%%%%%%%%%%%%%

\section{Color structure to four loops}
\label{sec-color}

Here we discuss the color dependence of  $\Gamma_{\rm cusp}$ to four loops.
It is best understood using results from non-Abelian eikonal exponentiation \cite{Gatheral:1983cz,Frenkel:1984pz}.

We start by setting up our conventions, following \cite{vanRitbergen:1998pn}. 
We consider a classical Lie-group with Lie-commutator
\begin{align}\label{eq:lie}
[ T^{a} ,T^{b} ] = i f^{abc} \, T^c \, ,
\end{align}
where the generators
\begin{align}\label{eq:generators}
T^{a}\, , \qquad a=1,\ldots,N_A
\end{align}
are taken in the fundamental representation. $f^{abc}$ are the structure constants of the Lie-algebra, and $N_{A}$ is the number of generators of the group.
The quadratic Casimir operators of the fundamental and adjoint representation of the Lie-algebra are
\begin{align}
[ T^{a} T^{a} ]_{ij} =& C_{F} \delta_{ij} \,, \qquad i,j=1,\ldots,N_F\\
f^{acd} f^{bcd} =& C_{A} \delta^{ab} \,,
\end{align}
respectively, where $N_F$ is the dimension of the fundamental represenation. The fundamental generators are normalized by ${\rm Tr}( T^{a} T^{b}) = T_{F} \delta^{ab}$.

The computation of color factors requires the evaluation of traces over products of generators. Up to three loops, at most six generators appear in the traces. Using the above equations, their result can be entirely expressed in terms of $C_{F}$ and $C_{A}$ (we normalize all color factors by ${\rm Tr}[1_{F}] = N_F$), e.g.~\cite{vanRitbergen:1998pn}
\begin{align}
{\rm Tr}(T^a T^b T^a T^b)/N_F =& C_F (C_F-C_A/2)\, , \\
{\rm Tr}(T^a T^b T^c T^a T^b T^c)/N_F = & C_F (C_F-C_A) (C_F-C_A/2)\, .
\end{align}
At four loops the trace over a product of eight generators can -- in general -- not be expressed solely in terms of $C_{F}$ and $C_{A}$, but higher group invariants are required. 
They can be expressed in terms of the following fully symmetrical tensors,
\begin{align}
d_{R}^{abcd} =& \frac{1}{6} {\rm Tr}[ T_R^a T_R^b T_R^c T_R^d + T_R^{a} T_R^b T_R^d T_R^c + T_R^a T_R^c T_R^b T_R^d  \nonumber \\
& \quad + T_R^a T_R^c T_R^d T_R^b + T_R^a T_R^d T_R^b T_R^c + T_R^a T_R^d T_R^c T_R^b ] \,.
\end{align}
Here $R$ can be either $F$ or $A$ for the fundamental and adjoint representation, respectively, with $ [T_F^a ]_{ij} \equiv  [T^a ]_{ij} $ and $ [T_A^a ]_{bc} = -i f^{abc}$.

Using the Lie-commutator one can show that up to terms proportional to powers of
$C_{F}$ and $C_{A}$, ${\rm Tr}(T^a T^b T^c T^d T^a T^b T^c T^d )$ is given by ${\rm Tr}(T^a T^b T^c T^d ) {\rm Tr}(T_A^a T_A^d T_A^c T_A^b ) $, which in turn is related to $d_{F}^{abcd} d_{A}^{abcd}$, see table 11 of~\cite{vanRitbergen:1998pn}. Explicitly, we have
\allowdisplaybreaks{
\begin{align}
{\rm Tr}(T^a T^b T^c T^d T^a T^b T^c T^d)/N_F &= \frac{d_{F}^{abcd} d_{A}^{abcd}}{N_F} + C_F \! \left[C_F^3-3 C_F^2 C_A +\frac{11}{4} C_F C_A^2-\frac{19}{24} C_A^3\right]  \label{eq:trace8}\, ,\\
\frac{d_{F}^{abcd} d_{A}^{abcd}}{N_F} &= {\rm Tr}(T^a T^b T^c T^d ) {\rm Tr}(T_A^a T_A^d T_A^c T_A^b )/N_F -\frac{1}{12} C_F C_A^3 \label{eq:quarticCasimir}\, .
\end{align}}
For a general Lie-group the traces over four generators in eq.~(\ref{eq:quarticCasimir}) cannot be expressed in terms of shorter traces which would lead to powers of $C_F$ and/or $C_A$. Hence we can consider $C_F$, $C_A$ and the quartic Casimir operator $d_{F}^{abcd} d_{A}^{abcd}/N_F$ as independent color structures at four loops, see ref.~\cite{vanRitbergen:1997va}.

{}From \cite{Gatheral:1983cz,Frenkel:1984pz} it follows that
Abelian-like terms containing powers of $C_{F}$ cancel 
in  $\Gamma_{\rm cusp}$, thanks to the logarithm in its definition, see eq.~(\ref{eq:defcuspAD}).
Moreover, an analysis of the possible color diagrams reveals that 
the result for $\Gamma_{\rm cusp}$ at one, two, and three 
loops is proportional to $C_{F}, C_{F} C_{A} , C_{F} C_{A}^2$, respectively.
At four loops, two structures appear, which we choose to be $C_{F} C_{A}^3$ and the 
quartic Casimir operator $d_{F}^{abcd} d_{A}^{abcd}/N_F$.

In summary, we have, to four loops
\begin{align}\label{color_general}
\log \langle W \rangle = g^2 C_{F} w_{1} + g^4 C_{F} C_{A} w_{2} + g^6 C_{F} C_{A}^2 w_{3} + g^8 \left[  C_{F} C_{A}^3 w_{4a} + \frac{d_{F}^{abcd} d_{A}^{abcd}}{N_F}  w_{4b} \right] \,,
\end{align}
where we have chosen the normalization $\langle W \rangle =1 + \cO(g^2)\,.$ We emphasize that
hitherto all relations are group-independent and apply to any of the classical Lie-groups.

The {\it webs} $w_{i}$ in (\ref{color_general}) correspond to linear combinations of Feynman diagrams.
The explicit expressions are easily obtained by the method of \cite{Gatheral:1983cz,Frenkel:1984pz}.
One advantage of this formulation is that one can directly compute the logarithm of the Wilson loop correlator,
and that each web only has an overall divergence\footnote{We tacitly assume that the intrinsic renormalization of the bare parameters of the Lagrangian has already been carried out.}. The latter is easy to remove, so that in practice one can define $\Gamma_{\rm cusp}$ in terms of finite integrals.

We now specialize the Lie-group to SU($N$), where all results can be explicitly written in terms of their
dependence on $N$. With the standard normalization for the fundamental generators, we have $N_F=N$ and
\begin{align}\label{colorN}
T_{F}=&\frac{1}{2} \,,\quad C_{A} =N \,,\quad C_{F} =\frac{N^2-1}{2 N} \,, \quad N_{A} = N^2-1 \,,\quad
 \frac{d_{F}^{abcd}d_{A}^{abcd} }{N_F} = \frac{(N^2-1) (N^2+6)}{48} \,.
\end{align}
Using eq. (\ref{colorN}), we make the dependence on $N$ of eq. (\ref{color_general}) manifest. As discussed above, we have exactly one color structure up to three loops, and two contributions at four loops, which
can now be distinguished thanks to their different dependence on $N$,
\begin{align}\label{general_color2}
\log \langle W \rangle = g^2  \frac{N^2-1}{2 N} \left[ w_{1} + g^2 N w_{2} + g^4 N^2 w_{3} + g^6 N^3 \left( w_{4a} +\frac{1}{24} w_{4b} \right) + g^6 N \frac{1}{4} w_{4b} \right] \,.
\end{align}
We see that in the large $N$ limit, keeping $\lambda=g^2 N$ fixed, only the contribution $g^6 N \frac{1}{4} w_{4b}$ disappears from the R.H.S. of eq. (\ref{general_color2}). In other words, to three loops, it is sufficient to know the planar
result for the Wilson loop in order to restore the full result in eq. (\ref{color_general}). At four loops, an additional
computation of the diagrams contributing to $w_{4b}$ is required.
In the remainder of this paper, we compute the 
non-planar contribution $w_{4b}$ in a recently-introduced scaling limit,
as well as the full planar result to four loops.

%%%%%%%%%%%%%%%%%%%%%%%%%%%%%%%%%%%%%%%%%%%%%%%%%%%%%%%%%%%%%%%%%%%%%%%%%%%%%%%%%%%%%%%%%%%%%%%%%%%%%%%%%%%%%%%%%%%%%%%%%%%%%%%%%%%%%%%%%%%%%%%%%%%%%
%%%%%%%%%%%%%%%%%%%%%%%%%%%%%%%%%%%%%%%%%%%%%%%%%%%%%%%%%%%%%%%%%%%%%%%%%%%%%%%%%%%%%%%%%%%%%%%%%%%%%%%%%%%%%%%%%%%%%%%%%%%%%%%%%%%%%%%%%%%%%%%%%%%%%

\section{Kinematics and integral functions}
\label{sec-kinematics}

As explained in the introduction, we will mainly be interested in the $\phi$
dependence of $\Gamma_{\rm cusp}$.
Here we discuss a convenient kinematical variable, and different physical regions. 
We also introduce a class of functions that we find appropriate to express the
answer in, and discuss the branch cut structure of the latter.

\subsection{Kinematical structure}

It is convenient to introduce a new variable $x=e^{i \phi}$, which in general is complex.
The computation we are considering is invariant under $\phi \to -\phi$. 
This corresponds to an inversion symmetry in $x$.

There are three different kinematical regions that we would like to discuss. 
It is useful to recall the relationship of $\Gamma_{\rm cusp}$ to IR
divergences of scattering processes involving massive particles, 
such as $e^+(p) \rightarrow \gamma^{*} e^+(q)$, which
have the same analytical structure.
(See e.g.\ refs. \cite{Czakon:2004wm,Anastasiou:2006hc}.)
With the on-shell conditions $p^2=q^2 =m^2$
(in the mostly-minus metric $+-\ldots -$), this process is naturally described 
using the variable $s / m^2$, where $s= (p-q)^2$.
It is related to $x$ via
\begin{align}
x = \frac{ \sqrt{1-4 m^2/s}-1}{\sqrt{1-4 m^2/s}+1} \,.
\end{align}
There we distinguish three kinematical regions, 
above threshold $s> 4 m^2$, 
below threshold, $0<s<4 m^2$, and finally the space-like region $s<0$.
They correspond to regions III, I , and II, respectively, that we now discuss
from the Wilson loop viewpoint.

\underline{Region I}: The first region corresponds to real $\phi$, $\phi \in [0,  \pi]$.
This means that the absolute value of $x$ is $1$.
In this case we have a cusp in Euclidean space, and 
$\Gamma_{\rm cusp}$ is real.
The two limiting cases are the following: for $\phi = 0$ the contour
is a straight line, and  $\Gamma_{\rm cusp}$ vanishes (for $\theta =0$).
The first correction $\sim \phi^2$ in this small angle limit
is known exactly in $\lambda$ and $N$ \cite{Correa:2012at,Fiol:2012sg}.
The opposite limit $\phi \to \pi$ is related to the quark-antiquark potential.
This limit is subtle and requires a resummation of certain diagrams, see 
\cite{Erickson:1999qv,Pineda:2007kz,Correa:2012nk,Bykov:2012sc,Stahlhofen:2012zx}.
One may also view the Wilson loop as the eikonal approximation to a 
form factor of massive quarks. In that case, this region corresponds to
the region below the two-particle threshold.

\underline{Region II}: We can analytically continue $\phi$ to Minkowskian angles.
In that case, $x$ is real and positive. Because of the inversion symmetry $x \to 1/x$,
it is sufficient to take $x\in [0,1]$.
The second endpoint, $x=1$, again corresponds  to the case of a straight line discussed above.
Near the endpoint $x\to0$, on the other hand, 
the cusp anomalous dimension diverges linearly
in  $\log(x)$, to all orders in the coupling constant \cite{Korchemsky:1987wg}. The coefficient of the
linear divergence is the well-studied light-like cusp anomalous dimension; the latter can
also be obtained from the anomalous dimension of high spin operators  \cite{Korchemsky:1988si,Korchemsky:1992xv,Alday:2007mf}. 
We may remark that the Wilson loop approach considered here is a very efficient way of computing this quantity.

\underline{Region III}:  Finally, we have the region above the threshold of creating two massive particles.
This region corresponds to $x$ being real and negative. As before, it suffices to take $x \in [-1,0]$, because
of the inversion symmetry in $x$. $\Gamma_{\rm cusp}$ has a branch cut along the negative real axis, and the $i 0$ prescription in the propagators implies that $x$ has a small imaginary part.
For the mostly minus Minkowski-space metric %$+-\ldots-$ 
that we are using, the position-space propagator
connecting two segments of the Wilson loop is proportional to $(s^2+t^2+s t(x+1/x) - i0)^{-1+\eps}$, where
the line parameters $s$ and $t$ are positive, and hence for $x \in [-1,0]$ we should add a small positive imaginary part to $x$.
In this region, $\Gamma_{\rm cusp}$ has an imaginary part.

\subsection{Harmonic polylogarithms and symbols}

What are the functions needed to describe $\Gamma_{\rm cusp}$?
Results at lower loop orders and for related scattering processes suggest that the class of functions we are seeking are the
harmonic polylogarithms (HPL)~\cite{Remiddi:1999ew}.
They are generalizations of ordinary polylogarithms,
and appear naturally e.g.\ within the differential equation technique to evaluate loop integrals, see e.g.\ \cite{Czakon:2004wm,Anastasiou:2006hc}.
They are also natural from the point of view of the singularity and branch cut structure described in the previous paragraph, with  $x=0,\pm 1$ being special points. They are defined iteratively by
\begin{eqnarray}\label{defHPL}
H_{a_1 , a_2 , \ldots, a_n }(x) = \int_0^x f_{a_1}(t) H_{a_2, \ldots, a_n }(t) \, dt \,, \quad \{a_1 , a_2 , \ldots a_n\} \neq \{0,0,\ldots 0\}\,,
\end{eqnarray}
where the integration kernels are defined as
\begin{align}
{f_{1}(x) = (1-x)^{-1}}\,,\qquad  {f_{0}(x) = x^{-1}}\,, \qquad {f_{-1}(x) = (1+x)^{-1}}\,.
\end{align}
The degree (or weight) $1$ functions needed to start the recursion are defined as
\begin{eqnarray}
H_{1}(x) &=& - \log(1-x) \,,\qquad H_{0}(x) = \log(x) \,, \qquad   H_{-1}(x) = \log(1+x) \,.
\end{eqnarray}
There is a special case when all indices are zero, $H_{\underbrace{0,\ldots,0}_{n}}(x) = \frac{1}{n!} \log^n (x)$.
The subscript of $H$ is called the weight vector. A common abbreviation is to replace occurrences
of $m$ zeros to the left of $\pm1$ by $\pm(m+1)$. For example, $H_{0,0,1,0,-1}(x) = H_{3,-2}(x)$.

HPLs have simple properties under certain argument transformations, and one can use
their algebraic properties to make their asymptotic behavior manifest.
We refer the interested reader to ref.~\cite{Remiddi:1999ew}. 
A very useful computer algebraic implementation has been given in refs.~\cite{Maitre:2005uu,Maitre:2007kp}.
For fast numerical evaluation, especially at complex arguments of the HPLs, we found the C++ 
implementation in GiNaC \cite{Bauer:2000cp} very helpful.

As we will describe in the section~\ref{sec-amplitude}, we compute the planar $\Gamma_{\rm cusp}$ from a massive scattering amplitude, where at each loop order a certain number of individual integrals appears. It turns out that each of these integrals can be expressed as a linear combination of HPLs of argument $x$ where in general all possible weight vectors appear at a given degree.
In the total result, however, we find the simplification that the result can be written in a compact form 
when using HPLs of argument $1-x^2$, and weight vectors with indices $0,1$ only! The latter property is also present in the four-loop non-planar correction in the scaling limit, and becomes manifest from the formulas in sections~\ref{sec-ladders} and~\ref{sec-result}.

In the context of the iterated integrals and differential equations studied in section \ref{sec-ladders}, the notion of 
the symbol of a transcendental function \cite{arXiv:math/0606419,2009arXiv0908.2238G,Goncharov:2010jf} is very useful.
It can be derived recursively for any function $f_w(x_1,\ldots,x_n)$ of weight $w$ whose total differential assumes the form
\begin{align}\label{eq:symbol1}
\displaystyle df_w = \sum\limits_{i} f_{i,w-1} \, d\log R_i \, ,
\end{align}
where the $f_{i,w-1}$ are of weight $w-1$ and the $R_i$ are algebraic functions. The symbol ${\cal S}(f_w)$ is then defined recursively via
\begin{align}\label{eq:symbol2}
\displaystyle {\cal S}(f_w) = \sum\limits_{i} {\cal S}(f_{i,w-1}) \, \otimes \, R_i \, ,
\end{align}
which involves a tensor product over the group of algebraic functions. We emphasize that eqns.~(\ref{eq:symbol1}) and~(\ref{eq:symbol2})
make the close connection between the `d-log'-representations (to be discussed in the next section) and the symbol of a function manifest.
We also note that symbols of the HPLs discussed above are built from the alphabet $\{ x, 1\pm x\}$. As a specific example, we have
\begin{align}
\displaystyle {\cal S}(H_n(x)) = {\cal S}({\rm{Li}}_n(x)) = - (1-x) \, \otimes \underbrace{x \otimes \cdots \otimes x}_{n-1 \; {\rm{ terms}}} \, .
\end{align}

%%%%%%%%%%%%%%%%%%%%%%%%%%%%%%%%%%%%%%%%%%%%%%%%%%%%%%%%%%%%%%%%%%%%%%%%%%%%%%%%%%%%%%%%%%%%%%%%%%%%%%%%%%%%%%%%%%%%%%%%%%%%%%%%%%%%%%%%%%%%%%%%%%%%%
%%%%%%%%%%%%%%%%%%%%%%%%%%%%%%%%%%%%%%%%%%%%%%%%%%%%%%%%%%%%%%%%%%%%%%%%%%%%%%%%%%%%%%%%%%%%%%%%%%%%%%%%%%%%%%%%%%%%%%%%%%%%%%%%%%%%%%%%%%%%%%%%%%%%%

\section{Iterated Wilson line integrals, and non-planar result in scaling limit}
\label{sec-ladders}

Here we discuss a general method for computing Wilson line integrals in position space.
We then apply it to the computation of the non-planar cusp anomalous dimension in the scaling limit.

\subsection{`d-log' forms for Wilson line integrals}
\label{lineintegrals}

\FIGURE[t]{
\includegraphics[width=0.60\textwidth]{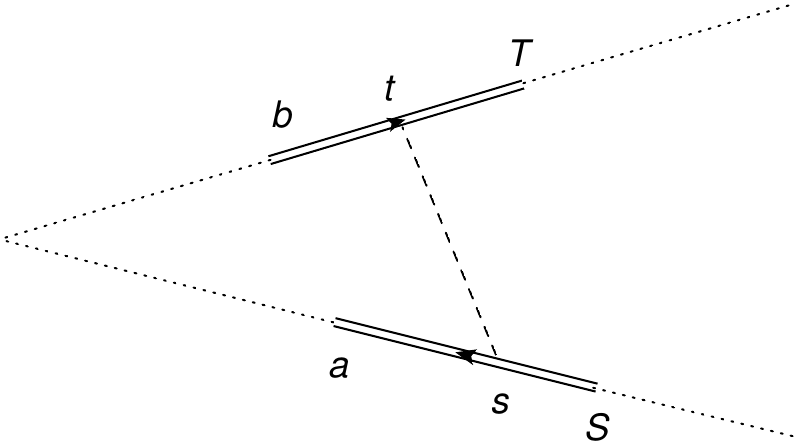}
\caption{Propagator-type integral discussed in the main text.}
\label{fig:example1loop}
}

In this section we elaborate on `d-log' forms for integrals, which were introduced in the context of scattering amplitudes in~\cite{CHTrento,ArkaniHamed:2012nw,Lipstein:2012vs}. As an instructive example, let us discuss the diagram shown in Fig~\ref{fig:example1loop}.
 The corresponding integral over the line parameters $s$ and $t$ can be written as
\begin{align}\label{dlogform}
\int_{\Lambda}  \frac{ds \wedge dt}{s^2 + t^2 + s t (x+1/x) } = \frac{x}{1-x^2} \, \int_{\Lambda} d \log ( s+ t x)  \wedge d \log (t + s x) \,,
\end{align}
where on the RHS we have dropped differentials involving $dx$ because they do not contribute to the integral,
and where the integration region ${\Lambda}$ is $s \in [a,S]$ and $t \in [b,T]$. 

What is gained from writing the integral in this way?
We see that a natural normalization factor, $x/(1-x^2)$, has been pulled out of the integral.
Together with trivial prefactors originating from the Feynman rules, this constitutes the normalization of the
diagram.
The remaining integral will give a (generalized) polylogarithmic function, which, in the present example, has degree $2$.
It depends on the variables $a,b,S,T$ and $x$,
\begin{align}
f(a,b,S,T,x)= \int_{\Lambda} d \log ( s+ t x)  \wedge d \log (t + s x) \,.
\end{align}
Integrals of this type satisfy simple differential equations, as we explain below.
Let us first focus on one of the two integration variables, say $s$, and rewrite the integral
in a more convenient form thanks to the identity \cite{CaronHuot:2011kk}
 \begin{align}\label{dlog_identity}
 d \log(s+\alpha)\wedge d\log(s+\beta) = d\log \frac{s+\alpha}{s+\beta} \wedge d \log (\alpha-\beta) \,.
 \end{align}
A simple generalization of this identity holds for $n$-forms.
 Then, we perform one integration at a time, in this case starting with the one over $s$.
 The main point is that one will always have an integral of the form
 \begin{align}\label{Simon_integral}
G(\alpha, \beta_{i}) := \int_{\Lambda_{y}} d \log(y+ \alpha ) F(y, \beta_{i})\,,
 \end{align}
 where $y$ is the integration variable, and $\alpha$ and $\beta_{i}$ are parameters, and $F$ is some function. 
 Then the algorithm outlined in appendix A of ref. \cite{CaronHuot:2011kk}
 can be used to determine the differential of $G$. It can be expressed in terms
 of quantities appearing in the differential of $F$.
 In our example, a short calculation gives
\begin{align}\label{differential-example}
d \, f(a,b,S,T,x)  =& d \log b \, \log \frac{(b+a x)(S+b x)}{(a+b x)(b+S x)} +d \log a \log \frac{(T+a x)(a+b x)}{(b+a x)(a+T x)} \nonumber \\
&+ d\log S \log \frac{(b+S x)(S+T x)}{(S+b x)(T+S x)} + d\log T \log \frac{(T+S x)(a+T x)}{(S+T x)(T+a x)}\nonumber\\
& + d\log x \log \frac{(T+a x)(S+b x)(b+S x)(a+T x)}{(b+a x)(a+b x)(T+S x)(S+T x)}\,. 
\end{align}
This equation determines $f$ up to an integration constant.
The latter can be determined, for example, from the boundary condition that $f$ vanishes at $a=S$.
Note that it is trivial to read off the symbol of $f$ from eq. (\ref{differential-example}).

In the present example, one can also directly integrate eq. (\ref{differential-example}).
The answer obtained can be written in terms of dilogarithms,
\begin{align}
f(a,b,S,T,x) =&
 {\rm Li}_{2}\left(-\frac{T}{S} x \right) - {\rm Li}_{2}\left( -\frac{T}{S} \frac{1}{x} \right)
 - {\rm Li}_{2}\left( -\frac{T}{a} x \right) + {\rm Li}_{2}\left( -\frac{T}{a} \frac{1}{x} \right) \nonumber\\
& - {\rm Li}_{2}\left( -\frac{b}{S} x \right)+ {\rm Li}_{2}\left( -\frac{b}{S} \frac{1}{x} \right)
 + {\rm Li}_{2}\left( -\frac{b}{a} x \right)- {\rm Li}_{2}\left( -\frac{b}{a} \frac{1}{x} \right) \,.
\end{align}
This agrees with ref. \cite{Makeenko:2008xr}.

In summary, we see that one can always compute the symbol of integrals that are of the type (\ref{dlogform}),
and for generalizations with more propagators, and polylogarithmic functions inserted into the integrand.
In particular, any ladder integral appearing in $\Gamma_{\rm cusp}$ can be computed in this way.
Preliminary results suggest that the generalization to graphs with interaction vertices 
is possible. For example, in ref. \cite{Henn:2012qz}, for two classes of diagrams the internal
integration associated to the interaction vertex could be computed analytically,
with the remaining integral of the form~(\ref{Simon_integral}).

We used this method to compute the non-planar ladder integrals appearing in  $\Gamma_{\rm cusp}$ to four loops.
In the next subsection, we discuss which integrals are required, and in the following subsection the results are reported.

\subsection{Non-planar contribution to scaling limit}

Here we compute the integrals contributing to $w_{4b}$ of eq. (\ref{color_general}) in the scaling limit.
Thanks to the scaling limit, we only need to keep ladder diagrams with four rungs between the two Wilson line segments.
Moreover, only diagrams containing the color factor $d_{F}^{abcd} d_{A}^{abcd}/N_F$ are required.\footnote{Recall that the color dependence of a general four-loop diagram can be expressed in terms of $C_{F}, C_{A}$, and ${d_{F}^{abcd} d_{A}^{abcd}}/N_F$. With our choice of color-basis in section 2, all terms with powers of $C_{F}$ higher than one cancel in $\log \langle W \rangle$, as per eq. (\ref{color_general}).}

\FIGURE[t]{
\includegraphics[width=0.95\textwidth]{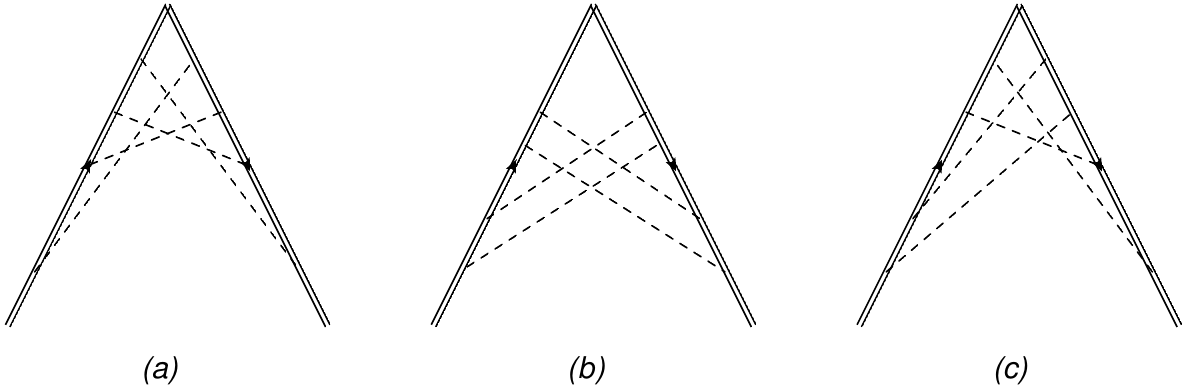}
\caption{All diagrams contributing to the quartic Casimir terms at four loops, in the scaling limit.}
\label{fig:np}
}
 It is easy to see that the only ladder type diagrams containing 
the quartic Casimir operator are the ones shown in Fig.~\ref{fig:np}.
Denoting their color factors by $\mathcal{C}_{i}$ and the integral functions by $\mathcal{I}_{i}$, with $i=a,b,c$
we have
\begin{align}
\log\langle W \rangle_{g^{8}} \sim \mathcal{C}_a \, \mathcal{I}_a +  \mathcal{C}_b \, \mathcal{I}_b +2 \, \mathcal{C}_c \, \mathcal{I}_c + \ldots\,,
\end{align}
where the dots represent other diagrams with color factors involving only $C_{F}$ and $C_{A}$, and the $2$ is a combinatorial factor, due to the fact that Fig.~\ref{fig:np}(c) also appears with the two Wilson lines interchanged.
The color factors of these diagrams contain a trace over eight
generators, e.g.\ ${\rm Tr}(T^a T^b T^c T^d T^a T^b T^c T^d )$ in Fig.~\ref{fig:np}(a),
and similarly for the other two diagrams.
Using the Lie commutator~(\ref{eq:lie}) one sees that the color factors of the three diagrams in Fig.~\ref{fig:np} are related. One finds
\begin{align}
\mathcal{C}_a =& {\rm Tr}(T^a T^b T^c T^d T^a T^b T^c T^d )/N_F \label{eq:colA} \, , \\
\mathcal{C}_c =& \, \mathcal{C}_a + \frac{1}{2} C_{F} C_{A} (C_{F}-C_{A}/2) (C_{F}-C_{A}) \,, \\
\mathcal{C}_b =& \, \mathcal{C}_a + \frac{1}{2} C_{F} C_{A} (C_{F}-C_{A}/2) (2 C_{F}-3/2 C_{A}) \label{eq:colC} \, ,
\end{align} 
where we normalize again all color factors by ${\rm Tr}[1_{F}] = N_F$.
From~(\ref{eq:colA}) --~(\ref{eq:colC}) we conclude that the three diagrams contribute equally to the color factor $d_{F}^{abcd} d_{A}^{abcd}/N_F$, see eq.~(\ref{eq:trace8}).
Taking this into account, we find that in the scaling limit the term proportional to 
$d_{F}^{abcd} d_{A}^{abcd}/N_F$ is given by
\begin{align}\label{resultwb}
w_{4b} \sim \mathcal{I}_a + \mathcal{I}_b + 2\, \mathcal{I}_c \,.
\end{align}

Let us discuss the definition of the integrals.
They are line integrals of the type considered in section \ref{lineintegrals}.
Here a comment on the regularization of the Wilson loop operator is due.
Naively, it has both infrared as well as ultraviolet divergences. 
We are interested in the ultraviolet divergences.
$\Gamma_{\rm cusp}$ is defined as the coefficient of the ultraviolet 
divergence. Since $\log \langle W \rangle$ only has an overall divergence,
it is easy to see how different regularization procedures are related.
The position space calculations above can be formulated e.g.\ in cut-off regularization
for both IR and UV divergences. 
Another possibility is to treat the integrals as in heavy-quark effective theory (HQET), with dimensional regularization.

In both cases, one can make the logarithmic divergence transparent by changing
variables. Let us denote the line parameters on the two lines by $s_{i}$ and
$t_{i}$, with $i=1,\ldots 4$, respectively. After rescaling all variables $s_{i} = \rho \tilde{s}_{i}, t_{i} = \rho \tilde{t}_{i}$,
with $\sum_{i=1}^{4} (\tilde{s}_i +\tilde{t}_{i}) =1$, the $\rho$ integral contains the divergence.
When working with cutoffs, this integral takes the form
\begin{align}
\int_{\Lambda_{\rm UV}}^{\Lambda_{\rm IR}} \frac{d\rho}{\rho} = \log \frac{\Lambda_{\rm IR}}{\Lambda_{\rm UV}} \,.
\end{align}
On the other hand, in HQET with dimensional regularization, one obtains
\begin{align}
\int_0^\infty \frac{d\rho}{\rho^{1-L \eps}} e^{-\rho}= \frac{1}{L \eps} + \cO(\eps^0) \,.
\end{align}
In both cases, the coefficient of the $\rho$ integral is the contribution to the cusp anomalous dimension that we wish to compute,
and it is given by convergent integrals.

For concreteness, let us choose the cutoff version of the calculation. In that
case, taking into account the discussion above and writing the integrals in d-log form as in section \ref{lineintegrals}, 
we have
\begin{align}\label{defitilde}
\mathcal{I}_{i} =  \log \frac{\Lambda_{\rm IR}}{\Lambda_{\rm UV}}  \, \left( \frac{x}{1-x^2} \right)^4 \, \tilde{\mathcal{I}}_{i} \,,\quad i=a,b,c\,.
\end{align}
Next, we can use the algorithm of section \ref{lineintegrals} to derive iterative differential equations
for $\tilde{\mathcal{I}}_{i}$. From these equations, we can immediately determine the symbol of these
functions as a corollary. We find that they are given by symbols composed from the alphabet $x , 1-x^2$.
This implies that they can be expressed in terms of a subset of the HPLs discussed in section \ref{sec-kinematics}, 
namely those with indices drawn from $0,1$, if we choose $x^2$ or $1-x^2$ as argument of the HPLs.

In order to determine the full functions from the differential equations, we have to complement them
by boundary conditions. The kinematical point $x=1$, or equivalently $\phi=0$, is a good boundary condition, 
where the Feynman integrals are expected to be regular. However, the prefactor $(x/(1-x^2))^4$ in (\ref{defitilde})
diverges in this limit, and hence the functions $\tilde{I}_{i}$ must have corresponding zeros.
We find it likely that a careful investigation of the iterative differential equations would reveal that this
boundary condition fixes all undetermined constants.
We found that simply using the condition of regularity of eq. (\ref{defitilde}) at $x=1$ determined most coefficients,
and we computed the remaining ones by considering asymptotic limits $x\to 0$, which we evaluated 
using standard Mellin-Barnes techniques. For more details on such methods, see section \ref{sec-amplitude}.
In this way, we found

\allowdisplaybreaks{
\begin{align}
 {\tilde{\mathcal I}}_{a}=&  -6\pi^2 H_{1, 1, 1, 2} + 
     48 H_{1, 1, 1, 4} - 8 \pi^2 H_{1, 1, 2, 1} + 
     64 H_{1, 1, 2, 3} + 64 H_{1, 1, 3, 2} \nonumber \\
     & - 6 \pi^2 H_{1, 2, 1, 1} + 48 H_{1, 2, 1, 3} + 
     48 H_{1, 2, 2, 2} - 10 \pi^2 H_{1, 1, 1, 1, 1} + 
     80 H_{1, 1, 1, 1, 3} \nonumber \\
     & + 80 H_{1, 1, 1, 2, 2} + 24 H_{1, 1, 1, 3, 1} 
     + 64 H_{1, 1, 2, 1, 2} + 
     32 H_{1, 1, 2, 2, 1} + 32 H_{1, 1, 3, 1, 1} \nonumber \\
     & + 48 H_{1, 2, 1, 1, 2} + 24 H_{1, 2, 1, 2, 1} + 
     24 H_{1, 2, 2, 1, 1} + 62 H_{1, 1, 1, 1, 1, 2} + 
     40 H_{1, 1, 1, 1, 2, 1} \nonumber \\
     & + 22 H_{1, 1, 1, 2, 1, 1} + 8 H_{1, 1, 2, 1, 1, 1} + 6 H_{1, 2, 1, 1, 1, 1} + 
     H_{1, 1, 1, 1, 1, 1, 1} \,, \label{eq:Itildea} \\
 {\tilde{\mathcal I}}_{b}=& -4 \pi^2 H_{1, 1, 1, 2} - 
    \frac{16}{3} \pi^2 H_{1, 1, 2, 1} + 
    16 H_{1, 1, 2, 3} + 32 H_{1, 1, 3, 2} - 
    4 \pi^2 H_{1, 2, 1, 1} \nonumber \\
    & + 16 H_{1, 2, 1, 3} + 16 H_{1, 2, 2, 2} - 
    \frac{20}{3} \pi^2 H_{1, 1, 1, 1, 1} + 
    16 H_{1, 1, 1, 1, 3} + 24 H_{1, 1, 1, 2, 2} \nonumber \\
    & + 24 H_{1, 1, 2, 1, 2} + 8 H_{1, 1, 2, 2, 1} + 
    16 H_{1, 1, 3, 1, 1} + 16 H_{1, 2, 1, 1, 2} + 
    8 H_{1, 2, 1, 2, 1} \nonumber \\
    &+ 8 H_{1, 2, 2, 1, 1} + 
    40 H_{1, 1, 1, 1, 1, 2} + 
    24 H_{1, 1, 1, 1, 2, 1} - 
    8 H_{1, 1, 2, 1, 1, 1} - 
    8 H_{1, 2, 1, 1, 1, 1} \nonumber \\
    & + 4 H_{1, 1, 1, 1, 1, 1, 1} + 
    48 \zeta_3 H_{1, 1, 1, 1}  \,,  \\
{\tilde{\mathcal I}}_{c}=& 
  + 4 \pi^2 H_{1, 1, 1, 2} - 12 H_{1, 1, 1, 4} + 
    \frac{16}{3} \pi^2 H_{1, 1, 2, 1} - 
    28 H_{1, 1, 2, 3} - 40 H_{1, 1, 3, 2} \nonumber \\
    &+ 4 \pi^2 H_{1, 2, 1, 1} - 
    24 H_{1, 2, 1, 3} - 24 H_{1, 2, 2, 2} + 
    \frac{20}{3} \pi^2 H_{1, 1, 1, 1, 1} - 
    32 H_{1, 1, 1, 1, 3}  \nonumber \\ 
   &- 38 H_{1, 1, 1, 2, 2} - 
    6 H_{1, 1, 1, 3, 1} - 34 H_{1, 1, 2, 1, 2} - 
    14 H_{1, 1, 2, 2, 1} - 20 H_{1, 1, 3, 1, 1} \nonumber \\
    & - 24 H_{1, 2, 1, 1, 2} - 12 H_{1, 2, 1, 2, 1} - 
    12 H_{1, 2, 2, 1, 1} - 
    38 H_{1, 1, 1, 1, 1, 2} - 
    22 H_{1, 1, 1, 1, 2, 1} \nonumber \\
    &- 4 H_{1, 1, 1, 2, 1, 1} + 
    2 H_{1, 1, 2, 1, 1, 1} + 
    2 H_{1, 2, 1, 1, 1, 1} + 
    2 H_{1, 1, 1, 1, 1, 1, 1} - 
    24 \zeta_3 H_{1, 1, 1, 1}  \label{eq:Itildec} \, .
\end{align}
}
Here we use the abbreviation $H_{w} = H_{w}(1-x^2)$.
Recalling eqs. (\ref{defitilde}) and (\ref{resultwb}), this determines $w_{4b}$ in the scaling limit.

We performed several consistency checks. First, using this algorithm, we reproduced the analytical result for the three-loop crossed ladder diagram computed in~\cite{Correa:2012nk}.
Moreover, we performed numerical checks on the above results
using the explicit line integral representation of the integrals.
Starting from the rescaled variables $\tilde{s}_{i}$ and $\tilde{t}_{i}$
above, we set
\begin{align}
\tilde{s}_{1} &= x_1 x_2 x_3 z \,, \quad \tilde{s}_{2} = x_1 x_2 z
\,,\quad \tilde{s}_{3} = x_1 z \,,\quad \tilde{s}_{4} = z \,,\\
\tilde{t}_{1} &= y_1 y_2 y_3 \bar{z} \,, \quad \tilde{t}_{2} = y_1 y_2
\bar{z} \,,\quad \tilde{t}_{3} = y_1 \bar{z} \,,\quad \tilde{t}_{4}
=\bar{z} \,,
\end{align}
where $\bar{z} :=1-z$, and with Jacobian $z^3 \bar{z}^3 x_1^2 x_2 y_1^2 y_2$.
Then we have, e.g.
\begin{align}
\tilde{I}_{a}=& \frac{(1-x^2)^4}{x^4} \int_0^1 dz \prod_{i=1}^{3} \left(
dx_i dy_i \right) z^3 \bar{z}^3 x_1^2 x_2 y_1^2 y_2 \times\nonumber \\
&\hspace{2 cm} \times  P(\tilde{s}_{1}, \tilde{t}_{4};x) P(\tilde{s}_{2},
\tilde{t}_{3};x)  P(\tilde{s}_{3}, \tilde{t}_{2};x) P(\tilde{s}_{4},
\tilde{t}_{1};x)\,,
\end{align}
where $P(s,t;x):=1/(s^2+t^2+s t(x+1/x))$. We used this formula (and corresponding ones for $\tilde{I}_{b}$ and $\tilde{I}_{c}$) to check (\ref{eq:Itildea}) -- (\ref{eq:Itildec}) numerically at the sub-per mille level for several values of $x$. Analytic checks can be done e.g.\ by switching to a Mellin-Barnes representation.

%%%%%%%%%%%%%%%%%%%%%%%%%%%%%%%%%%%%%%%%%%%%%%%%%%%%%%%%%%%%%%%%%%%%%%%%%%%%%%%%%%%%%%%%%%%%%%%%%%%%%%%%%%%%%%%%%%%%%%%%%%%%%%%%%%%%%%%%%%%%%%%%%%%%%
%%%%%%%%%%%%%%%%%%%%%%%%%%%%%%%%%%%%%%%%%%%%%%%%%%%%%%%%%%%%%%%%%%%%%%%%%%%%%%%%%%%%%%%%%%%%%%%%%%%%%%%%%%%%%%%%%%%%%%%%%%%%%%%%%%%%%%%%%%%%%%%%%%%%%

\section{Planar calculation from massive scattering amplitude}
\label{sec-amplitude}

It was shown in ref. \cite{Henn:2010bk} that the Regge limit $s/m^2 \gg 1$ of the planar
Coulomb branch amplitude $M(s/m^2,t/m^2)$ is governed by the cusp anomalous dimension.
This connection was used in \cite{Correa:2012nk} to compute the three-loop value of $\Gamma_{\rm cusp}$.

The advantage of this approach is that an expression for the integrand of $M(s/m^2,t/m^2)$ is
already known. It is in the form of a small number of scalar integrals, each of which results from
many Feynman diagrams. 
This simple integrand was obtained by using generalized unitarity, in conjunction with (extended)
dual conformal symmetry \cite{Bern:2006ew,Henn:2010ir}.
It has been pointed out that the limit relating the amplitude and $\Gamma_{\rm cusp}$ also works
at the level of the integrand \cite{Henn:2012qz}. This implies that one can obtain an efficient integral representation
for $\Gamma_{\rm cusp}$ in this way.

Here we wish to extend the work of \cite{Correa:2012nk} to four loops and determine the planar part of $\Gamma_{\rm cusp}$ from
the four-loop scattering amplitude. As a starting point, convenient Mellin-Barnes representations for the eight contributing integrals are
available from~\cite{Henn:2010ir}. The strategy of our calculation is the following:
First, we use generalized cuts to determine the power of $\xi$ to which each scattering amplitude integral contributes.
Details of this procedure can be found in appendix~\ref{app:integrals}.
Next, based on experience from lower loop orders and the structure observed for the ladder diagrams~\cite{Henn:2012qz}, we make an ansatz subject to
certain assumptions, which we will specify below.
This reduces the calculation to the determination of a certain number of undetermined coefficients.
In order to determine the latter, we analyze both the Mellin-Barnes representations and the ansatz in various limits, such
as $x\to 0$ and $x\to 1$. In this way, we obtain (more than) enough algebraic equations to determine all coefficients. Moreover,
this provides consistency checks for the ansatz.

\subsection{Assumptions and classification of HPLs}

An analysis of the cuts of the integrals contributing to the scattering amplitude 
suggests that the four-loop result for $\theta=0$, where we have $\xi = (1-x)/(1+x)$, has the structure
\begin{align}\label{structure-four-loops}
\Gamma_{\rm cusp} |_{\lambda^4/(8 \pi^2)^4} = \sum_{r=1}^{4} \left( \frac{1-x}{1+x} \right)^r \Gamma^{(4;r)}(x)  + \cO(1/N^2) \,,
\end{align}
where the $\Gamma^{(4;r)}(x)$ are certain transcendental functions.
What can we assume about their structure? Looking at the results up to three loops we may make a number of observations.
\begin{itemize}
\item all results for $\Gamma_{\rm cusp}$ can be written in terms of harmonic polylogarithms
\item the degree of transcendentality of the functions involved is uniformly $(2L-1)$, where $L$ is the loop order
\item the subset of HPLs with indices $0,1$ only and argument $1-x^2$ is sufficient to describe the answer. In terms of the symbol, this means that only letters $x, 1-x^2$ are required.
\end{itemize}
In the case of ladder integrals, the first two items can be proved, and the third item is true at least up to six loops~\cite{Henn:2012qz}.
As we showed in section~\ref{sec-ladders}, it is also true in the non-planar case.
We find it reasonable to assume that these properties hold true for the full result at four loops.

If this assumption is correct, the calculation is reduced to the determination of the precise linear combination of the
allowed functions. 
Our starting point will be all functions $H_{w}(1-x^2)$ of weight seven, where the weight vector $w$ is 
build from entries $0,1$. We also allow transcendental constants such as $\zeta_i, \zeta_i \zeta_j$, possibly multiplying lower degree functions to construct a term of total degree seven.

We can restrict and classify these functions according to their symmetry properties.
In fact, $\Gamma_{\rm cusp}$ has to be symmetric under the inversion $x \to 1/x$. This follows from the definition
$\cos \phi = 1/2 (x+1/x)$. As a result,  $\Gamma^{(4;r)}(x)$ has to be odd/even under this symmetry for $r$ odd/even.
In this way, we find $51$ even and $50$ odd functions under this symmetry.

Another simple condition we can impose is that $\Gamma^{(4;r)}(x)$ should have at least $r$ zeros as $x \to 1$.
The reason is that it is multiplied by the factor $\xi^r$, which for $\theta=\pi/2$ has a degree $r$ pole at $x = 1$. 
But $x=1$ corresponds to the straight line case, and this should be finite for each integral contributing to $\Gamma_{\rm cusp}$.
We will also verify this behavior by expanding integrals near $x\to 1$.

In order to further restrict the number of functions, we make two additional observations about the results at lower loops $L\le 3$, eqs. (\ref{result1loopxi1})-(\ref{result3loopxi3}).
Inspecting them one sees that in these cases the number of $0$'s in the weight vector of the functions $\Gamma^{(L;r)}(x)$ 
is always smaller than $r$. 
 We will assume this to hold true also at four loops.
Moreover, $\Gamma^{(L;r)}(x)$ vanishes for $L>1$ at $x=-1$.\footnote{In this case one has to rewrite $\Gamma^{(L;r)}(x)$ in terms of HPLs of argument $x$. We will discuss the analytic continuation of $\Gamma_{\rm cusp}$ in the next section and in appendix~\ref{app:analyticcont}.} This is required in order to obtain the correct leading order behavior at $x\to-1$, which corresponds to the quark antiquark potential limit. This limit will be discussed in more detail at the end of this section.

Imposing all these conditions our ansatz becomes, in summary,
\begin{itemize}
\item $12$ functions for $\Gamma^{(4;2)}(x)$, of degree $7$, indices $0,1$, even under $x\to 1/x$, at most one $0$ entry in weight vector, and vanishing as $(1-x)^2$ as $x \to1$ and as $(1+x)^1$ as $x \to -1$.
\item $21$ functions for $\Gamma^{(4;3)}(x)$, of degree $7$, indices $0,1$, odd under $x\to 1/x$, at most two $0$ entries in weight vector, and vanishing as $(1-x)^3$ as $x \to1$ and as $(1+x)^1$ as $x \to -1$.
\end{itemize}
The terms $\Gamma^{(4;1)}$ and $\Gamma^{(4;4)}$ were already computed in refs.~\cite{Correa:2012at} and~\cite{Henn:2012qz}, respectively. 

\subsection{Asymptotic limit of Mellin-Barnes integrals}

Let us now explain how to determine the coefficients of the ansatz. By means of the Mathematica packages
{\tt MB.m}~\cite{Czakon:2005rk} and {\tt MBasymptotics.m}~\cite{MBasymptotics} we perform the asymptotic expansions of
the Mellin-Barnes representations, first about the point $x=0$. The expansion parameter appears in the form $x^p \, \log^q(x)$,
and each of these terms is accompanied by one or several transcendental constants. For $p>0$ these constants are in general not
of homogeneous weight, but the highest transcendentality is always $7-q.$\footnote{Except for $q=6$, where it is zero.}
We determine these constants analytically for $p=0,\ldots,6$ and $q=3,\ldots,7$. For $p=0$ we also include $q=2$. It is interesting
to note that at most two-dimensional Mellin-Barnes integrals are required for this calculation at $q>2$, and three-dimensional ones at $q=2$.
After we computed all relevant constants in this way, we perform the series expansion of our ansatz to the respective orders in $x$
and $\log(x)$, and solve the resulting algebraic equations for the unknown coefficients appearing in our ansatz. As an illustrating
example, take
\be\label{eq:illustrating}
\left[\frac{22}{9} + \frac{2\pi^2}{3}\right] \, x \, \log^5(x) \stackrel{!}{=} 
\left[\left(-\frac{8}{27} \, a_1+\frac{28}{9} \, a_2\right) + \pi^2 \, \left(-\frac{1}{6} \, a_1 + \frac{1}{3} \, a_2 \right)\right] \, x \, \log^5(x) \; ,
\ee 
where the LHS stems from the solution of the Mellin-Barnes integrals at a particular power in $x$
and $\log(x)$, and the RHS stands for the expansion of the ansatz to the same order. Assuming $1$ and $\pi^2$
to be linearly independent we obtain two {\emph{algebraic}} equations, yielding $a_1=-3$ and $a_2=1/2$.

Since some coefficients in our ansatz appear only at powers $q=1$ or $q=0$, the above procedure does yield most, but not all
coefficients. In order to determine the remaining ones we expand the Mellin-Barnes representations about the point $x=1$.
In this limit the expansion is purely of the form $(x-1)^s$, without any logarithms. We include all terms with $s \le 4$
and determine the coefficients in the same way as above. However, this time we have to solve Mellin-Barnes integrals that are
up to seven-dimensional.

We emphasize that the number of equations this procedure yields is much larger than the number of undetermined coefficients
in our ansatz, such that our system of equations is largely overconstrained. As a rule of thumb we have about twice as
many equations compared to the number of coefficients. This property is of utmost importance since otherwise potential
inconsistencies in our ansatz could not be revealed.

We also do numerical checks, but only after the application of {\tt MBasymptotics.m}, i.e.\
we check numerically all analogues of the $x$-independent part of the LHS of eq.~(\ref{eq:illustrating}).
Performing numerical checks on the unexpanded expressions is not well suited here since by construction
the integral and the ansatz agree to high powers in $x$ and $(x-1)$. Hence the ansatz
obtained in this way will agree very well numerically with the integral we are computing, even if the ansatz 
was incomplete.

Last but not least we have the algebraic cross-check that the final answer does only diverge linearly in $\log(x)$
as $x \to 0$, see eq.~(\ref{eq:lightlike1}). This cross-check is non-trivial since it connects different powers of $\xi$,
each of which diverges with the seventh power of $\log(x)$.

\subsection{Planar result to four loops}
\label{sec-result}

We are now in the position to present the results for the cusp anomalous dimension up to four loops in the planar limit. We have
\begin{align}\label{result-four-loop}
\Gamma_{\rm cusp}(x, \theta=0,\lambda,N) = \sum_{L=1}^{4}  \left( \frac{\lambda}{8 \pi^2}\right)^L \sum_{r=1}^{L} \left( \frac{1-x}{1+x} \right)^r \Gamma^{(L;r)}(x)  + \cO(\lambda^5,1/N^2) \,,
\end{align}
where
\begin{align}
\Gamma^{(1;1)} =& \frac{1}{2} H_{1} \,, \label{result1loopxi1}\\
\Gamma^{(2;1)} =&-\frac{1}{4} H_{1,1,1}-\frac{1}{6} \pi ^2 H_1 \,, \label{result2loopxi1}\\
\Gamma^{(2;2)} =&\frac{1}{2} H_{1,2} +\frac{1}{4} H_{1,1,1} \label{result2loopxi2}
\end{align}
at one and two loops \cite{Korchemsky:1987wg,
Makeenko:2006ds,
Kidonakis:2009ev,
Drukker:2011za},
\allowdisplaybreaks{
\begin{align}
\Gamma^{(3;1)} =&\frac{1}{4} \pi ^2 H_{1,1,1}+\frac{5}{8} H_{1,1,1,1,1}+\frac{\pi ^4}{12} H_1 \,, \label{result3loopxi1}\\
\Gamma^{(3;2)} =&-\frac{3}{2} \zeta_3 H_{1,1}-\frac{1}{6} \pi ^2 H_{1,2}-\frac{1}{3} \pi ^2 H_{2,1}-\frac{1}{4} \pi ^2
   H_{1,1,1}-H_{1,1,1,2}-\frac{3}{4} H_{1,2,1,1}\nonumber \\
   &-H_{2,1,1,1}-\frac{11}{8} H_{1,1,1,1,1} \,, \label{result3loopxi2} \\
\Gamma^{(3;3)} =& H_{1,1,3}+H_{1,2,2}+H_{1,1,1,2}+\frac{1}{2} H_{1,1,2,1}+\frac{1}{2} H_{1,2,1,1}+\frac{3}{4} H_{1,1,1,1,1} \label{result3loopxi3}
\end{align}
}
at three loops \cite{Correa:2012nk},
and
\allowdisplaybreaks{
\begin{align}
\Gamma^{(4;1)} =& -\frac{1}{5} \pi ^4 H_{1,1,1}-\pi ^2 H_{1,1,1,1,1}-\frac{7}{2} H_{1,1,1,1,1,1,1}-\frac{2}{45} \pi ^6 H_1 \,,  \label{result4loopxi1} \\
\Gamma^{(4;2)} =&
\frac{45}{4} \zeta_5 H_{1,1}+\frac{2}{3} \pi ^2 \zeta_3 H_{1,1}+5 \zeta_3 H_{1,1,1,1}+\frac{1}{12} \pi ^4
   H_{1,2}+\frac{5}{18} \pi ^4 H_{2,1}+\frac{13}{72} \pi ^4 H_{1,1,1} \nonumber \\
   & +\pi ^2 H_{1,1,1,2}+\pi ^2 H_{1,1,2,1}+\frac{3}{4} \pi
   ^2 H_{1,2,1,1}+\frac{5}{3} \pi ^2 H_{2,1,1,1}+\frac{53}{24} \pi ^2 H_{1,1,1,1,1} \nonumber \\
   & +5 H_{1,1,1,1,1,2}+\frac{7}{2}
   H_{1,1,1,1,2,1}+\frac{9}{2} H_{1,1,1,2,1,1}+3 H_{1,1,2,1,1,1}+\frac{25}{8} H_{1,2,1,1,1,1} \nonumber \\
   & +\frac{25}{4}
   H_{2,1,1,1,1,1}+\frac{203}{16} H_{1,1,1,1,1,1,1}
   \,,  \label{result4loopxi2} \\
\Gamma^{(4;3)} =& 
-3 \zeta_3 H_{1,1,2}-4 \zeta_3 H_{1,2,1}-3 \zeta_3 H_{2,1,1}-5 \zeta_3 H_{1,1,1,1}-\frac{1}{120} \pi ^4
   H_{1,1,1}-\frac{2}{3} \pi ^2 H_{1,1,3} \nonumber \\ 
& -\frac{2}{3} \pi ^2 H_{1,2,2}-\pi ^2 H_{1,3,1}-\pi ^2 H_{2,1,2}-\pi ^2
   H_{2,2,1}-\frac{7}{6} \pi ^2 H_{1,1,1,2}-\frac{4}{3} \pi ^2 H_{1,1,2,1} \nonumber \\
 &  -\frac{5}{6} \pi ^2 H_{1,2,1,1}-\pi ^2
   H_{2,1,1,1}-\frac{29}{24} \pi ^2 H_{1,1,1,1,1}-5 H_{1,1,1,1,3}-\frac{7}{2} H_{1,1,1,2,2}
   \nonumber \\
   &
   -3 H_{1,1,2,1,2}-2
   H_{1,1,2,2,1}-3 H_{1,1,3,1,1}-5 H_{1,2,1,1,2}-4 H_{1,2,1,2,1}-\frac{9}{2} H_{1,2,2,1,1}
   \nonumber \\
   &
   -3 H_{1,3,1,1,1}-5 H_{2,1,1,1,2}-6
   H_{2,1,1,2,1}-5 H_{2,1,2,1,1}-3 H_{2,2,1,1,1}-\frac{43}{4} H_{1,1,1,1,1,2}
   \nonumber \\
   &
   -\frac{17}{2} H_{1,1,1,1,2,1}-8
   H_{1,1,1,2,1,1}-7 H_{1,1,2,1,1,1}-\frac{33}{4} H_{1,2,1,1,1,1}-\frac{19}{2} H_{2,1,1,1,1,1}\nonumber \\
   &
   -\frac{239}{16}
   H_{1,1,1,1,1,1,1} \,,  \label{result4loopxi3} \\
\Gamma^{(4;4)} =& 3 H_{1,1,1,4}+4 H_{1,1,2,3}+4 H_{1,1,3,2}+3 H_{1,2,1,3}+3 H_{1,2,2,2}+5 H_{1,1,1,1,3}+5 H_{1,1,1,2,2}
\nonumber \\
   &+\frac{3}{2}
   H_{1,1,1,3,1}+4 H_{1,1,2,1,2}+2 H_{1,1,2,2,1}+2 H_{1,1,3,1,1}+3 H_{1,2,1,1,2}+\frac{3}{2} H_{1,2,1,2,1}
   \nonumber \\
   &
   +\frac{3}{2}
   H_{1,2,2,1,1}+\frac{23}{4} H_{1,1,1,1,1,2}+5 H_{1,1,1,1,2,1}+4 H_{1,1,1,2,1,1}+3 H_{1,1,2,1,1,1}\nonumber \\
   &+\frac{9}{4}
   H_{1,2,1,1,1,1}+\frac{23}{4} H_{1,1,1,1,1,1,1}  \label{result4loopxi4}
\end{align}}
at four loops. As mentioned above, the terms $\Gamma^{(4;1)}$ and $\Gamma^{(4;4)}$ were already computed
in refs. \cite{Correa:2012at} and \cite{Henn:2012qz}, respectively. The remaining terms $\Gamma^{(4;2)}$ and $\Gamma^{(4;3)}$ are new.
We derived them analytically, subject to the assumptions discussed in the previous section.

In the above equations $H_{w} := H_{w}(1-x^2)$, and $x=e^{i \phi}$.
The perturbative results given in this section and in section~\ref{sec-ladders} can be straightforwardly evaluated
numerically in the region II, i.e.\ $0<x<1$. Other regions can be reached by analytical continuation,
respecting the branch cut properties discussed in section~\ref{sec-kinematics}. We collect the relevant formulas
in appendix~\ref{app:analyticcont}.

A curious feature of the result up to four loops, already remarked upon in \cite{Henn:2012qz}, is that once the result is written in terms of HPLs with argument $1-x^2$ as above, all HPLs in $(-1)^{(r+L)} \Gamma^{(L;r)}$ come with non-negative coefficients.

In Figures~\ref{fig:region2} and~\ref{fig:region1} we plot the cusp anomalous dimension in Regions II and I, respectively.
From the plots one can see the properties discussed below and in section \ref{sec-kinematics}.

\FIGURE[t]{
\includegraphics[width=0.7\textwidth]{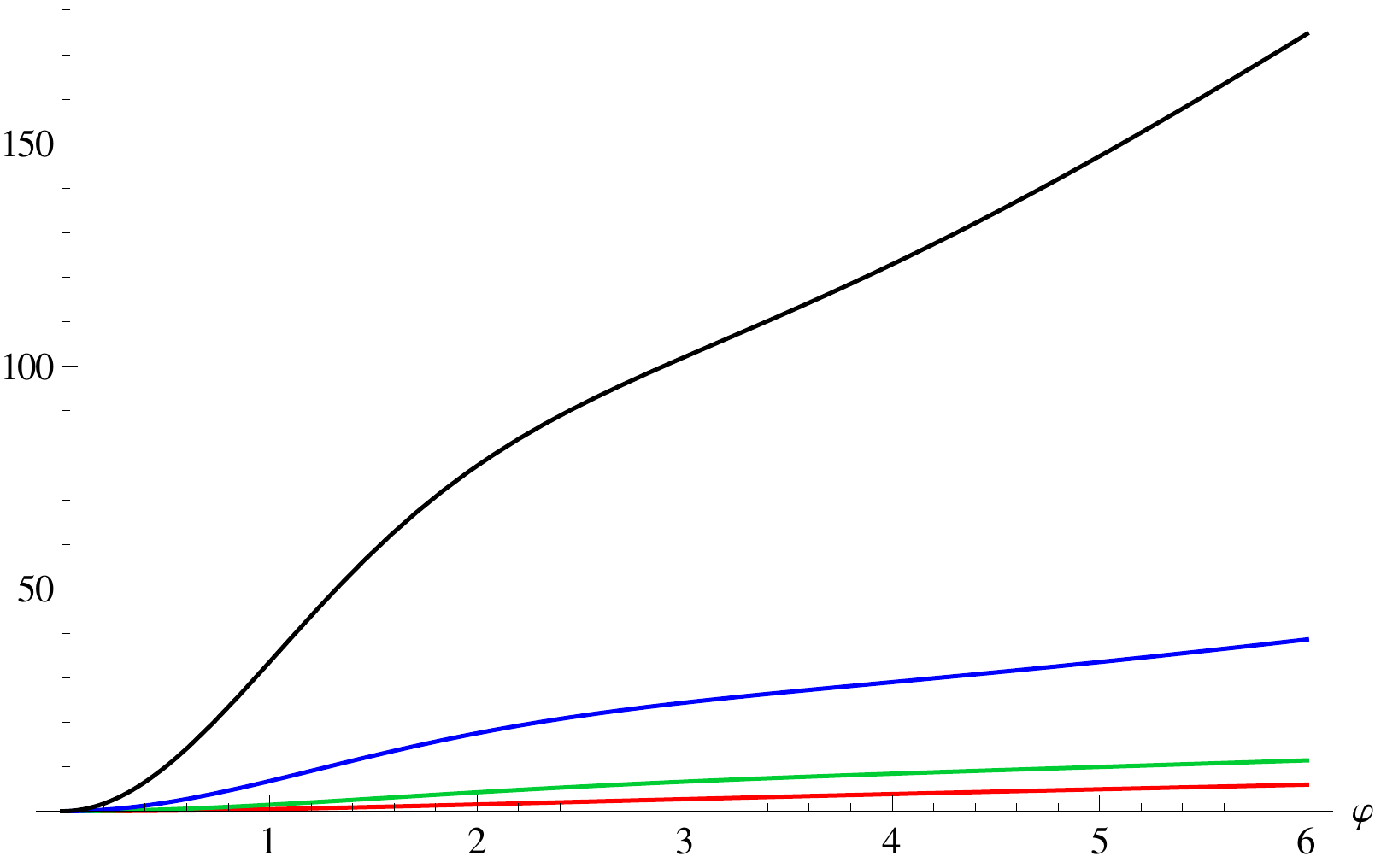}
\begin{picture}(0,0)%
\setlength{\unitlength}{1pt}%
\put(10,58){{\raisebox{45mm}[0mm][0mm]{%
           \makebox[-240mm]{\scalebox{0.95}{$ (-1)^{L+1} \Gamma_{\rm cusp}^{(L)} $}}}}}%
\end{picture}%
\caption{Plot of $(-1)^{L+1} \Gamma_{\rm cusp}^{(L)}$ as a function of $\varphi = -\log(x)$ with $x \in [0,1]$, i.e.\ Region II. From bottom to top the plot shows $L=1,2,3,4$.
The small and large $\varphi$ behavior is known to all loop orders:
For small $\varphi$ the first term is quadratic, with the coefficient given by the Bremsstrahlung function.
At large $\varphi$, $\Gamma_{\rm cusp}$ grows linearly, with the coefficient determined by the light-like cusp anomalous dimension.}
\label{fig:region2}}

We now consider various limits of $\Gamma_{\rm cusp}$. First, we can use the above results to analytically compute the light-like cusp anomalous dimension.
It is obtained by taking the limit $x \to 0$, where $\Gamma_{\rm cusp}$ diverges logarithmically,
\begin{align}
\lim_{x \to 0} \Gamma_{\rm cusp} = -\frac{1}{2} \log (x) \, \Gamma^{\infty} +{ \mathcal{G}}_{0} +\cO(x) \,. \label{eq:lightlike1}
\end{align}
We find
\begin{align}
\Gamma^{\infty} = 2 \left( \frac{\lambda}{8 \pi^2} \right) - \frac{\pi^2}{3} \left( \frac{\lambda}{8 \pi^2} \right)^2 + \frac{11 \pi^4}{90} \left( \frac{\lambda}{8 \pi^2} \right)^3 + \left( -2 \zeta_3^2-\frac{73 \pi ^6}{1260}  \right) \left( \frac{\lambda}{8 \pi^2} \right)^4 + \cO(\lambda^5) \,. \label{eq:lightlike2}
\end{align}
This agrees with previous numerical results at four loops~\cite{Bern:2005iz,Bern:2006ew,Cachazo:2006az,Henn:2010ir}, and with the spin chain prediction
from ref.~\cite{Beisert:2006ez}.
The behaviour~(\ref{eq:lightlike1}) can also be seen from Fig.~\ref{fig:region2}, where the curves grow linearly for large values of $\varphi=-\log(x)$.
For ${ \mathcal{G}}_{0} $ we find
\begin{align}
{ \mathcal{G}}_{0} = &
 -\zeta_3 \left( \frac{\lambda}{8 \pi^2} \right)^2 
+ \left( \frac{9 \zeta_5}{2}-\frac{\pi ^2 \zeta_3}{6}\right)  \left( \frac{\lambda}{8 \pi^2} \right)^3 
+ \left( \frac{\pi ^4 \zeta_3}{10}+\frac{11 \pi ^2 \zeta_5}{12}-\frac{85 \zeta_7}{4} \right)  \left( \frac{\lambda}{8 \pi^2} \right)^4 \nnb \\
& + \cO(\lambda^5,\lambda^4/N^2)
\end{align}
${ \mathcal{G}}_{0}$  is related to the collinear anomalous dimension for mass-regulated scattering amplitudes~\cite{Henn:2010bk}.
Unlike $\Gamma^{\infty}$, this quantity depends on the regularization scheme and takes a different value in dimensional regularization~\cite{Cachazo:2007ad}.

\FIGURE[t]{
\includegraphics[width=0.7\textwidth]{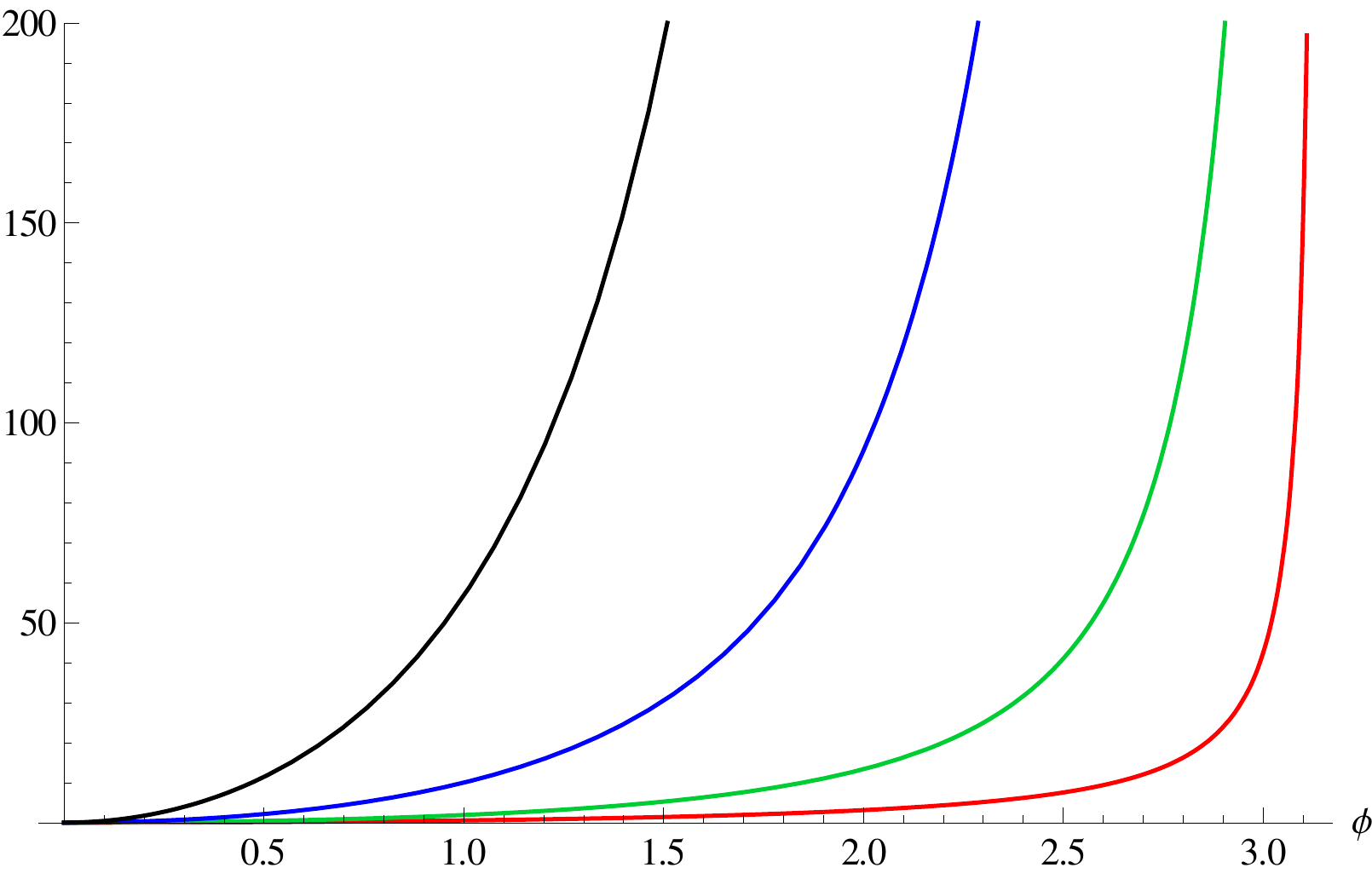}
\begin{picture}(0,0)%
\setlength{\unitlength}{1pt}%
\put(10,58){{\raisebox{45mm}[0mm][0mm]{%
           \makebox[-245mm]{\scalebox{0.95}{$ (-1)^L \Gamma_{\rm cusp}^{(L)} $}}}}}%
\end{picture}%
\caption{The functions $(-1)^L \Gamma_{\rm cusp}^{(L)}$ in the interval $\phi \in [0,\pi]$, i.e.\ Region I. From bottom to top the plot shows $L=1,2,3,4$.}
\label{fig:region1}}

Next we consider $\phi \to 0$, corresponding to $x=e^{i\phi} \to 1$, where we find
\be\label{eq:smallphi}
\dps \Gamma_{\rm cusp} = \phi^2 \left[-\frac{1}{2} \left( \frac{\lambda}{8 \pi^2} \right) + \frac{\pi^2}{6} \left( \frac{\lambda}{8 \pi^2} \right)^2
-\frac{\pi^4}{12} \left( \frac{\lambda}{8 \pi^2} \right)^3 + \frac{2\pi^6}{45} \left( \frac{\lambda}{8 \pi^2} \right)^4\right] + {\cal O}(\phi^3) \, .
\ee
This behaviour can be seen from Figs.~\ref{fig:region2} and~\ref{fig:region1}, where all curves start quadratically from the respective origin.
The expansion in~(\ref{eq:smallphi}) is in perfect agreement with the four-loop expansion of the exact result in~\cite{Correa:2012at,Fiol:2012sg}.

The third limit to consider is $x \to -1$, which we parameterize by $x=e^{i\phi}$, $\phi=\pi-\delta$, and $\delta \to 0$. We find \allowdisplaybreaks{
\begin{align}
\xi \, \Gamma^{(1;1)} =& \, -\frac{2\pi}{\delta}+ {\cal O}(\delta^0) \; , \\
\xi \, \Gamma^{(2;1)} =& \; {\cal O}(\delta^0) \; , \\
\xi^2 \, \Gamma^{(2;2)} =& \, -\frac{8\pi}{\delta} \, L_\delta+{\cal O}(\delta^0) \; , \\
\xi \, \Gamma^{(3;1)} =& \; {\cal O}(\delta^0) \; , \\
\xi^2 \, \Gamma^{(3;2)} =& \, -\frac{\pi}{\delta} \left[\frac{16\pi^2}{3} \, L_\delta + 36\zeta_3 +\frac{16\pi^2}{3}\right] + {\cal O}(\delta^0) \; , \\
\xi^3 \, \Gamma^{(3;3)} =& \, -\frac{8\pi^4}{3\delta^2} -\frac{\pi}{\delta}\left[16 \, L^2_\delta + 16 \, L_\delta -4\pi^2 -24\right] + {\cal O}(\delta^0) \; , \\
\xi \, \Gamma^{(4;1)} =& \; {\cal O}(\delta^0) \; , \\
\xi^2 \, \Gamma^{(4;2)} =& \, -\frac{\pi}{\delta} \left[ 32 \pi^2 \zeta_3 - 190 \zeta_5 \right]+ {\cal O}(\delta^0) \; , \\
\xi^3 \, \Gamma^{(4;3)} =& \, -\frac{16\pi^6}{9\delta^2} - \frac{\pi}{\delta}\left[\frac{64\pi^2}{3} \, L^2_\delta + \left(96 \zeta_3 + \frac{272\pi^2}{3}\right) L_\delta -\frac{8\pi^4}{3}+48\zeta_3-\frac{208\pi^2}{3}\right] \nnb \\
&+ {\cal O}(\delta^0) \; , \\
\xi^4 \, \Gamma^{(4;4)} =& \, 32 \zeta_3 \, \frac{\pi^3}{\delta^3} -  \frac{\pi^2}{\delta^2} \left[\frac{64\pi^2}{3} \, L^2_\delta + 64 \zeta_3+\frac{16\pi^2}{3}\right] - \frac{\pi}{\delta} \left[\frac{64}{3} \, L^3_\delta + 64 \, L^2_\delta\right. \nnb \\
& \left.+\left(-\frac{112\pi^2}{3} - 32\right) L_\delta +\frac{32}{3} \, \pi^2\zeta_3 +96\zeta_3+\frac{16\pi^2}{9}-\frac{512}{3}\right] + {\cal O}(\delta^0) \; ,
\end{align}
}
with $L_\delta = \ln(2\delta/e)$. Figure~\ref{fig:region1} shows the divergences as $\phi \to \pi$.

The limit $x\to -1$ is related to the quark antiquark potential.
This limit is subtle. Due to ultrasoft effects, a resummation is required.
This is done by matching fixed order calculations against an effective field theory calculation.
In the context of $\cN=4$ SYM, this was discussed in ref. \cite{Pineda:2007kz}, and more recently in \cite{Stahlhofen:2012zx}.

We close this section with two remarks. First, note that for $\theta=0$, we have $\xi = (1-x)/(1+x)$.
It is natural to expect that the full $\theta$ dependence can be obtained by replacing $((1-x)/(1+x))^r$ in eq.~(\ref{result-four-loop})
by $\xi^r$, see section~\ref{sec-intro}.

Our second remark concerns the regularization scheme dependence. The above method assumed a supersymmetric regularization
scheme, and therefore we expect our result for $\Gamma_{\rm cusp}$ to be valid in that scheme.
The transition to other schemes, such as $\overline{{\rm MS}}$, is discussed in ref. \cite{Belitsky:2003ys}, and has been explicitly worked out
there to two loops.

%%%%%%%%%%%%%%%%%%%%%%%%%%%%%%%%%%%%%%%%%%%%%%%%%%%%%%%%%%%%%%%%%%%%%%%%%%%%%%%%%%%%%%%%%%%%%%%%%%%%%%%%%%%%%%%%%%%%%%%%%%%%%%%%%%%%%%%%%%%%%%%%%%%%%
%%%%%%%%%%%%%%%%%%%%%%%%%%%%%%%%%%%%%%%%%%%%%%%%%%%%%%%%%%%%%%%%%%%%%%%%%%%%%%%%%%%%%%%%%%%%%%%%%%%%%%%%%%%%%%%%%%%%%%%%%%%%%%%%%%%%%%%%%%%%%%%%%%%%%

\section{Comparison to strong coupling via AdS/CFT}
\label{sec-strong}

Can we compare the fixed order perturbative results of section \ref{sec-result} to the 
results available at strong coupling via the AdS/CFT correspondence?
The authors of ref.~\cite{Kotikov:2003fb} proposed such a procedure in the
case of the light-like cusp anomalous dimension.
They combined perturbative data with the string
theory insight that the strong coupling expansion takes the form
\begin{align}\label{expansion-strong}
\Gamma_{\rm cusp} = c  \sqrt{\lambda} + \ldots \,,
\end{align}
where $c$ is negative, and we work in the planar limit.
In order to incorporate this behavior they
proposed the following ansatz $f(\lambda)$ for $\Gamma_{\rm cusp}$ 
\begin{align}\label{KLVansatz}
\lambda^n = \sum_{r=n}^{2n} C_{r} \left[ f(\lambda) \right]^r \, ,
\end{align}
where $n$ is connected to the loop order $L$ via $n=L-1$. The constants $C_{r}$ can be fixed using perturbative information.
Of course, one can also use strong coupling data, as in \cite{Bern:2006ew}, in order to
gain insights on weak coupling. Here we will use the perturbative two-,
three-, and four-loop results for $\Gamma_{\rm cusp}$ in order to determine the coefficient of $\sqrt{\lambda}$
at strong coupling.

Let us give more details about this ansatz in the simplest case, i.e.\ $n=1$.
Here the extrapolation is based on the two-loop perturbative information, i.e.\
\begin{align}
\Gamma_{\rm cusp} = v_1 \lambda + v_2 \lambda^2 + \cO(\lambda^3) \,.
\end{align}
Then one can determine the coefficients in eq. (\ref{KLVansatz}) to be
$C_{1} = 1/v_1, C_{2} = -v_2/v_1^3$.
The latter equation then implies that the ansatz for the interpolation function is
\begin{align}
f(\lambda) = -\frac{ v_1^2}{2 v_2 } \left[ -1 + \sqrt{1- 4 \lambda  \frac{v_2}{v_{1}}} \right] \,.
\end{align}
which at strong coupling gives $- \sqrt{\lambda} \sqrt{- \frac{v_1^3}{v_2}} + \ldots $.
This procedure can be generalized to higher loops, where eq.~(\ref{KLVansatz}) implies that one has to
solve equations of higher degree, which can be done numerically.

We then compare this extrapolation to results obtained via the AdS/CFT correspondence.
At strong coupling, i.e.\ $\lambda \gg 1$, Wilson loops are described by minimal surfaces \cite{Maldacena:1998im}.
We have the expansion (\ref{expansion-strong}), and focus on the case $\theta=0$.
The result for $c$ for region I, i.e.\ real angles, is available through implicit equations involving Elliptic integrals
from~\cite{Drukker:1999zq}. In the case of region II, we can use the formulas of~\cite{Kruczenski:2002fb}.
The first subleading coefficient in the expansion (\ref{expansion-strong}) is also known~\cite{Forini:2010ek}.
Let us discuss the results of the comparison for the two regions in turn.

\underline{Extrapolation for region I}: We used the ansatz of eq. (\ref{KLVansatz}) in order to extrapolate the strong-coupling
coefficient $c$ in eq. (\ref{expansion-strong}) from the knowledge of the four-loop data.
%We present some sample values in table \ref{extrapolation}.
We found that the extrapolations based on our four loop results and the 
$\sqrt{\lambda}$ behavior give a leading order strong coupling answer
that agrees to within $2$ per cent for the range of $\phi \in [0.1, 2.5]$.
However, for $\phi>2.5$ the relative error grows significantly. This is not 
surprising since there one approaches the quark-antiquark limit $\phi \to \pi$.

\underline{Extrapolation for region II}: Here we find very good agreement between the extrapolation based on the four-loop perturbative data,
and the strong coupling answer. 
It is interesting to note that the relative error to the strong coupling value goes down from 
approximately $25\%$, $3\%$ and under $1.6\%$, when using two-loop, three-loop and four-loop data
as input, respectively. 
It is also remarkable that this relative error stays small for all data points analyzed in the 
interval $x\in [0,1]$, despite the fact that the leading coefficient (in front of $\sqrt{\lambda}$) at strong 
coupling varies by several orders of magnitude.

Let us comment on the radius of convergence of the expansions.
It is known for the $x \to 1$ and $x \to 0$ limits, respectively.
The former is described by the Bremsstrahlung function \cite{Correa:2012at}, 
whose perturbative series has a radius of convergence of $\lambda_c \approx 14.7 $.
The latter is governed by the light-like cusp anomalous dimension, 
where the radius of convergence \cite{Beisert:2006ez} is $\lambda_c = \pi^2$.

%%%%%%%%%%%%%%%%%%%%%%%%%%%%%%%%%%%%%%%%%%%%%%%%%%%%%%%%%%%%%%%%%%%%%%%%%%%%%%%%%%%%%%%%%%%%%%%%%%%%%%%%%%%%%%%%%%%%%%%%%%%%%%%%%%%%%%%%%%%%%%%%%%%%%
%%%%%%%%%%%%%%%%%%%%%%%%%%%%%%%%%%%%%%%%%%%%%%%%%%%%%%%%%%%%%%%%%%%%%%%%%%%%%%%%%%%%%%%%%%%%%%%%%%%%%%%%%%%%%%%%%%%%%%%%%%%%%%%%%%%%%%%%%%%%%%%%%%%%%

\section{Conclusion and Outlook}
\label{sec-discussion}

We computed the velocity-dependent cusp anomalous dimension
in maximally supersymmetric Yang-Mills theory to four-loop order.
The result can be expressed in terms of harmonic polylogarithms
of degree seven, with argument $1-x^2$ and non-negative indices only.
We determine the non-planar correction at four loops in the scaling limit,
which involves quartic Casimir invariants as color factors. The method of
`d-log'-representations for iterated Wilson line integrals turns out to
be extremely powerful for this purpose. It allows one to compute the symbol of such functions.
If the symbols correspond to a known class of functions, HPLs in our case,
one can integrate back using boundary conditions.

Moreover, we determine the full planar four-loop result from massive scattering amplitudes, 
where we use asymptotic expansions of Mellin-Barnes
integrals to analytically pin down the coefficients of a well-motivated ansatz.
Our analytical result gives the correct values of the four-loop light-like cusp anomalous dimension
that was previously calculated only numerically~\cite{Bern:2005iz,Bern:2006ew,Cachazo:2006az,Henn:2010ir}.

We also compare our perturbative result to strong coupling, and find that
our extrapolation agrees to better than two per cent with the corresponding
string theory result, over a wide range of parameters.

Taken together, the only pieces missing to obtain the full -- planar and non-planar --
result of the velocity-dependent cusp anomalous dimension to four loops
are the non-planar terms proportional to $\xi^2$ and $\xi^3$. The light-like
limit of the non-planar cusp anomalous dimension is also envisaged in~\cite{Boels:2012ew} by means
of the on-shell form factor. However, we emphasize that
the present approach allows to obtain the full $x$-dependence, and not just the
light-like limit.

The results we have derived here shed light on the structure of 
the planar four-particle amplitude on the Coulomb branch of $\cN=4$ super Yang-Mills. 
The latter is an infrared finite function $M(s/m^2,t/m^2)$.
Kinematically it is very similar to light-by-light scattering via massive particles.
It is an interesting open question what class of two-variable functions describe such processes
beyond the one-loop order.
The integrals we have computed determine the asymptotic limit of this amplitude as $s/m^2 \gg 1$.

There are several generalizations to the `d-log'-approach discussed 
in section \ref{sec-ladders}. The first generalization concerns the Wilson loop contour.
While we have focused on a contour formed by two segments in this
paper, it is clear that  the technique applies equally to contours formed by $n$ segments meeting in a point.  
This is relevant for the description of infrared divergences of massive scattering amplitudes at the
non-planar level, see e.g. \cite{Ferroglia:2009ii,Mitov:2009sv,Chien:2011wz}.
We also wish to emphasize that massless results can be obtained as a corollary.

Another obvious generalization of the applicability of this technique has to do with
the regularization.  On physical grounds, at least in principle, one can always choose combinations of 
diagrams that only have a superficial UV divergence.
For such quantities, one can easily switch between regulators.
Our method is very naturally formulated in a cut-off scheme, however it is equally possible
to use dimensional regularization. This is straightforward for integrals 
that only have a superficial UV divergence. For other integrals, one first has to 
identify the integration regions that lead to divergences and perform subtractions.

Finally, whereas we focussed in the present paper on scalar and gluon exchanges, preliminary results suggest
that the generalization to graphs with interaction vertices is possible~\cite{Henn:2012qz}.

%%%%%%%%%%%%%%%%%%%%%%%%%%%%%%%%%%%%%%%%%%%%%%%%%%%%%%%%%%%%%%%%%%%%%%%%%%%%%%%%%%%%%%%%%%%%%%%%%%%%%%%%%%%%%%%%%%%%%%%%%%%%%%%%%%%%%%%%%%%%%%%%%%%%%
%%%%%%%%%%%%%%%%%%%%%%%%%%%%%%%%%%%%%%%%%%%%%%%%%%%%%%%%%%%%%%%%%%%%%%%%%%%%%%%%%%%%%%%%%%%%%%%%%%%%%%%%%%%%%%%%%%%%%%%%%%%%%%%%%%%%%%%%%%%%%%%%%%%%%

\section*{Acknowledgments}
It is a pleasure to thank Dirk Seidel and Simon Caron-Huot for very useful correspondence. We thank R.~Britto, L.~Dixon, E.~Gardi, G.~Korchemsky, L.~Magnea, J.~Maldacena, M.~Neubert, D.~Simmons-Duffins, and A.~Sever for useful discussions.
J.M.H. is  grateful to the ECT* Trento for hospitality during part of this work,
and wishes to thank the organizers of  ``Periods and Motives'', Paris 2012, where part of this work was presented, for their invitation.
J.M.H. was supported in part by the Department of Energy grant DE-FG02-90ER40542, and the IAS AMIAS fund.
The work of T.H. was partially supported by the Helmholtz Alliance ``Physics at the Terascale''. 

\appendix

%%%%%%%%%%%%%%%%%%%%%%%%%%%%%%%%%%%%%%%%%%%%%%%%%%%%%%%%%%%%%%%%%%%%%%%%%%%%%%%%%%%%%%%%%%%%%%%%%%%%%%%%%%%%%%%%%%%%%%%%%%%%%%%%%%%%%%%%%%%%%%%%%%%%%
%%%%%%%%%%%%%%%%%%%%%%%%%%%%%%%%%%%%%%%%%%%%%%%%%%%%%%%%%%%%%%%%%%%%%%%%%%%%%%%%%%%%%%%%%%%%%%%%%%%%%%%%%%%%%%%%%%%%%%%%%%%%%%%%%%%%%%%%%%%%%%%%%%%%%

\section{Four-loop integrals and generalized cuts / leading singularities}
\label{app:integrals}

The velocity-dependent cusp anomalous dimension $\Gamma_{\rm cusp}(\phi)$
can be obtained from the Regge limit of massive amplitudes in $\cN=4$ super Yang-Mills~\cite{Henn:2010bk,Henn:2010ir}.
At the four-loop level, there are eight contributing diagrams, which are depicted in Figure~1 of~\cite{Henn:2010ir}.
The corresponding amplitude is given by eq.~(2.8) of that reference.
In the Regge limit $s\to\infty$ the logarithm of the amplitude is given by\footnote{Note that refs. \cite{Henn:2010bk,Henn:2010ir} use different metric conventions.}
\begin{align}
\log{\cal M} \stackrel{s\to\infty}{\longrightarrow} \log(-m^2/s) \, \Gamma_{\rm cusp}(-m^2/t) \, ,
\end{align}
where $t$ is related to $x$ via $-m^2/t=x/(1-x)^2$.

We now study the systematics of the Regge limit at the four-loop level~\cite{Henn:2010ir} by considering
the integrals contributing to the four-loop amplitude.
We expect them to have the general structure
\begin{align}\label{assumption}
I = I_{0} \times \tilde{I}\,,
\end{align}
where $I_{0}$ is an algebraic normalization factor, and
$\tilde{I}$ is a function having degree of transcendentality eight.
In the literature, such functions are sometimes referred to as pure functions.

Generalized cuts or leading singularities are useful in order to test
whether (\ref{assumption}) holds, and to determine the normalization
factor $I_{0}$.

For example, for the massive box integral at one loop, normalized by $s t$, we have
\begin{align}
I_{0} \sim  1/\sqrt{1- 4 m^2/s-4 m^2/t} \,.
\end{align}
Notice that in the Regge limit, this factor becomes proportional to $\xi$.

Likewise, we computed the maximal cuts of all integrals up to four loops.
The result is consistent with eq. (\ref{assumption}), and we find the following
behavior of the prefactors as $s\to \infty$
(the superscript ``$r$'' means that the integral is rotated, i.e.\ $s \leftrightarrow t$)
\allowdisplaybreaks{\begin{align}
& I_{4a} \sim \xi,\qquad I^{r}_{4a} \sim \xi^4 \nnb \\
& I_{4b} \sim \xi^2,\qquad I^{r}_{4b} \sim \xi \nnb \\
& I_{4c} \sim \xi,\qquad I^{r}_{4c} \sim \xi^3 \nnb \\
& I_{4d} \sim \xi,\qquad I^{r}_{4d} \sim \xi^3 \nnb \\
& I_{4e} \sim \xi,\qquad I^{r}_{4e} \sim \xi^2 \nnb \\
& I_{4f} \sim \xi,\qquad I^{r}_{4f} \sim \xi^2 \nnb \\
& I_{4d2} \sim \xi,\qquad I^{r}_{4d2} \sim \xi^2 \nnb \\
& I_{4f2} \sim \xi^2,\qquad I^{r}_{4f2} \sim \xi^2 \, .
\end{align}
}
Notice that due to exponentiation, the maximal power of Regge logarithms that  a given
integral has is bounded by the power of $\xi$. Comparing to Appendix A of~\cite{Henn:2010ir},
we find that this is in agreement with the above $\xi$ dependence.

We see that we can classify the contribution of the integrals to $\Gamma_{\rm cusp}$ 
according to which power of $\xi$ that they are normalized by. This is a very useful feature,
as it allows to compute the contributions to different powers of $\xi$ independently.

%%%%%%%%%%%%%%%%%%%%%%%%%%%%%%%%%%%%%%%%%%%%%%%%%%%%%%%%%%%%%%%%%%%%%%%%%%%%%%%%%%%%%%%%%%%%%%%%%%%%%%%%%%%%%%%%%%%%%%%%%%%%%%%%%%%%%%%%%%%%%%%%%%%%%
%%%%%%%%%%%%%%%%%%%%%%%%%%%%%%%%%%%%%%%%%%%%%%%%%%%%%%%%%%%%%%%%%%%%%%%%%%%%%%%%%%%%%%%%%%%%%%%%%%%%%%%%%%%%%%%%%%%%%%%%%%%%%%%%%%%%%%%%%%%%%%%%%%%%%

\section{Analytic continuation of $\Gamma_{\rm cusp}$}
\label{app:analyticcont}

In order to analytically continue $\Gamma_{\rm cusp}$ to regions~I and~III it is sufficient to apply
the argument transformation $1-x^2 \to x^2$ to the HPLs in section~\ref{sec-result}
and subsequently extract the logarithms explicitly. This gives
\begin{align}
\Gamma^{(1;1)} =& -\log(x) \, , \\
\Gamma^{(2;1)} =& +\frac{1}{3} \log^3(x) + \frac{\pi^2}{3} \log(x) \,, \\
\Gamma^{(2;2)} =& -\frac{1}{3} \log^3(x) - \frac{\pi^2}{6} \log(x) - \log(x) \, H_{2}(x^2) + H_{3}(x^2) - \zeta_3 
\end{align}
at one and two loops,
\allowdisplaybreaks{
\begin{align}
\Gamma^{(3;1)} =&- \frac{1}{6} \log^5(x) - \frac{\pi^2}{3} \log^3(x) - \frac{\pi^4}{6} \log(x) \,, \\
\Gamma^{(3;2)} =&  -6 H_{5}(x^2)+\frac{2}{3} \log^4(x) H_{1}(x^2)-\frac{1}{3} \log ^3(x) H_{2}(x^2)+\frac{2\pi ^2}{3}
    \log ^2(x) H_{1}(x^2) \nonumber \\
   & -\log ^2(x) H_{3}(x^2)-\frac{\pi ^2}{3}  \log (x) H_{2}(x^2)+\frac{9}{2} \log (x) H_{4}(x^2)+\zeta_3 \log^2(x) \nonumber \\
   &  +\frac{11}{30} \log ^5(x)+\frac{5\pi ^2}{9}  \log ^3(x)+\frac{5\pi ^4}{36} \log (x)+6 \zeta_5  \,,  \\
\Gamma^{(3;3)} =& -2 \zeta_3 H_{2}(x^2)+\frac{\pi ^2}{6} 
   H_{3}(x^2)+3 H_{5}(x^2)+2
   H_{2,3}(x^2)+3 H_{3,2}(x^2)+3
   H_{4,1}(x^2) \nonumber \\
   & -\frac{2}{3} \log ^3(x) H_{2}(x^2)+\log ^2(x)
   H_{3}(x^2)-\frac{\pi ^2}{3}  \log (x) H_{2}(x^2)-2 \log
   (x) H_{4}(x^2) \nonumber \\
   & -2 \log (x) H_{2,2}(x^2)-2 \log (x)
   H_{3,1}(x^2)-\zeta_3 \log ^2(x)-\frac{1}{5} \log ^5(x)-\frac{2\pi ^2}{9}  \log^3(x) \nonumber \\
   &-\frac{\pi ^4}{30}  \log (x)-\frac{3}{2} \zeta_5-\frac{\pi ^2}{6} \zeta_3
\end{align}
}
at three loops, and
\allowdisplaybreaks{
\begin{align}
\Gamma^{(4;1)} =& +\frac{4}{45}\log ^7(x)+\frac{4\pi ^2}{15}
    \log ^5(x)+\frac{4\pi ^4}{15}  \log ^3(x)+\frac{4\pi ^6}{45}  \log (x) \,,   \\
\Gamma^{(4;2)} =&-\frac{29}{90} \log ^7(x)-\frac{5}{9}
   H_{1}(x^2) \log ^6(x)+\frac{5}{6} H_{2}(x^2) \log
   ^5(x)-\frac{73}{90} \pi ^2 \log ^5(x) \nonumber \\
   & -\frac{10}{9} \pi ^2 H_{1}(x^2) \log
   ^4(x)-2 H_{3}(x^2) \log ^4(x)-\zeta_3 \log ^4(x)+\frac{11}{9} \pi ^2
   H_{2}(x^2) \log ^3(x) \nonumber \\
   & +\frac{11}{6} H_{4}(x^2) \log
   ^3(x)-\frac{161}{270} \pi ^4 \log ^3(x)-\frac{5}{9} \pi ^4 H_{1}(x^2) \log
   ^2(x)-\frac{7}{3} \pi ^2 H_{3}(x^2) \log ^2(x) \nonumber \\
   & +\frac{11}{2}
   H_{5}(x^2) \log ^2(x)-\frac{11}{2} \zeta_5 \log ^2(x)-\frac{2}{3} \pi ^2 \zeta_3 \log ^2(x)+\frac{7}{18} \pi ^4 H_{2}(x^2) \log (x) \nonumber \\
   &+\frac{17}{6} \pi ^2 H_{4}(x^2) \log (x)-\frac{95}{4} H_{6}(x^2) \log
   (x)-\frac{353 \pi ^6}{3780} \log (x)-\frac{1}{9} \pi ^4 H_{3}(x^2) \nonumber \\
   &-\frac{2}{3} \pi ^2 H_{5}(x^2)+35 H_{7}(x^2)-35 \zeta_7+\frac{2 \pi
   ^2 \zeta_5}{3}+\frac{\pi ^4 \zeta_3}{9} \,, \\
\Gamma^{(4;3)} =& +\frac{239}{630} \log ^7(x)+\frac{38}{45}
   H_{1}(x^2) \log ^6(x)-\frac{1}{3} H_{2}(x^2) \log
   ^5(x)+\frac{4}{5} \pi ^2 \log ^5(x) \nonumber \\
   &+\frac{11}{9} \pi ^2 H_{1}(x^2) \log ^4(x)+2
   H_{1,2}(x^2) \log ^4(x)+2 H_{2,1}(x^2) \log ^4(x)+2 \zeta_3 \log ^4(x) \nonumber \\
   &-\frac{2}{9} \pi ^2 H_{2}(x^2) \log ^3(x)+3
   H_{4}(x^2) \log ^3(x)-\frac{4}{3} H_{1,3}(x^2) \log
   ^3(x)-2 H_{2,2}(x^2) \log ^3(x) \nonumber \\
   &-4 H_{3,1}(x^2) \log
   ^3(x)+\frac{4}{3} H_{1}(x^2) \zeta_3 \log ^3(x)+\frac{47}{108} \pi ^4 \log
   ^3(x)+\frac{17}{45} \pi ^4 H_{1}(x^2) \log ^2(x) \nonumber \\
   &+\frac{2}{3} \pi ^2
   H_{3}(x^2) \log ^2(x)-10 H_{5}(x^2) \log ^2(x)+2 \pi ^2
   H_{1,2}(x^2) \log ^2(x) +9 H_{4}(x^2) \zeta_3\nonumber \\
   &+2 \pi ^2 H_{2,1}(x^2) \log ^2(x)-2
   H_{2,3}(x^2) \log ^2(x)-2 H_{3,2}(x^2) \log ^2(x)+8 \zeta_5 \log ^2(x) \nonumber \\
   &+2 H_{2}(x^2) \zeta_3 \log ^2(x)+\frac{4}{3} \pi ^2 \zeta_3
   \log ^2(x)+\frac{1}{90} \pi ^4 H_{2}(x^2) \log (x)-\frac{7}{6} \pi ^2
   H_{4}(x^2) \log (x) \nonumber \\
   &+20 H_{6}(x^2) \log (x)-2 \pi ^2
   H_{1,3}(x^2) \log (x)-2 H_{1,5}(x^2) \log (x)-\frac{8}{3}
   \pi ^2 H_{2,2}(x^2) \log (x) \nonumber \\
   &+7 H_{2,4}(x^2) \log
   (x)-\frac{8}{3} \pi ^2 H_{3,1}(x^2) \log (x)+10
   H_{3,3}(x^2) \log (x)+16 H_{4,2}(x^2) \log (x) \nonumber \\
   &+22 H_{5,1}(x^2) \log (x)+2 H_{1}(x^2) \zeta_5 \log (x)+2
   \zeta_3^2 \log (x)+2 \pi ^2 H_{1}(x^2) \zeta_3 \log (x) \nonumber \\
   &-4 H_{3}(x^2) \zeta_3 \log (x)+\frac{38}{945} \pi ^6 \log (x)-\frac{1}{45} \pi ^4
   H_{3}(x^2)-32 H_{7}(x^2)+\frac{2}{3} \pi ^2
   H_{2,3}(x^2) \nonumber \\
   &-10 H_{2,5}(x^2)+\pi ^2
   H_{3,2}(x^2)-13 H_{3,4}(x^2)+\pi ^2
   H_{4,1}(x^2)-18 H_{4,3}(x^2)-28 H_{5,2}(x^2) \nonumber \\
   &-40 H_{6,1}(x^2)+16 \zeta_7+10
   H_{2}(x^2) \zeta_5+\frac{\pi ^2}{2} \zeta_5-\frac{2}{3} \pi ^2
   H_{2}(x^2) \zeta_3+\frac{\pi ^4
   \zeta_3}{45}
 \,,  \\
\Gamma^{(4;4)} =& -\frac{46}{315} \log ^7(x)-\frac{3}{5} H_{2}(x^2) \log ^5(x)-\frac{23}{90}
   \pi ^2 \log ^5(x)+H_{3}(x^2) \log ^4(x)-\zeta_3 \log ^4(x) \nonumber \\
   &-\frac{2}{3} \pi ^2 H_{2}(x^2) \log ^3(x)-\frac{7}{3} H_{4}(x^2) \log ^3(x)-2
   H_{2,2}(x^2) \log ^3(x)-\frac{8}{3} H_{3,1}(x^2) \log
   ^3(x) \nonumber \\
   &-\frac{19}{180} \pi ^4 \log ^3(x)+\frac{2}{3} \pi ^2 H_{3}(x^2) \log
   ^2(x)+4 H_{5}(x^2) \log ^2(x)+3 H_{2,3}(x^2) \log ^2(x) \nonumber \\
   &+6 H_{3,2}(x^2) \log ^2(x)+9 H_{4,1}(x^2) \log
   ^2(x)-\frac{5}{2} \zeta_5 \log ^2(x)-3 H_{2}(x^2) \zeta_3 \log
   ^2(x) \nonumber \\
   &-\frac{2}{3} \pi ^2 \zeta_3 \log ^2(x)-\frac{1}{10} \pi ^4 H_{2}(x^2)
   \log (x)-\frac{2}{3} \pi ^2 H_{4}(x^2) \log (x)-4
   H_{6}(x^2) \log (x) \nonumber \\
   &-\pi ^2 H_{2,2}(x^2) \log (x)-6
   H_{2,4}(x^2) \log (x)-\frac{4}{3} \pi ^2 H_{3,1}(x^2) \log
   (x)-10 H_{3,3}(x^2) \log (x) \nonumber \\
   &-16 H_{4,2}(x^2) \log (x)-22
   H_{5,1}(x^2) \log (x)-6 H_{2,2,2}(x^2) \log (x)-6
   H_{2,3,1}(x^2) \log (x) \nonumber \\
   &-8 H_{3,1,2}(x^2) \log (x)-8
   H_{3,2,1}(x^2) \log (x)-6 H_{4,1,1}(x^2) \log (x)-\zeta_3^2 \log (x) \nonumber \\
   &+2 H_{3}(x^2) \zeta_3 \log (x)-\frac{17 \pi ^6}{2520}\log(x)+\frac{1}{30} \pi ^4 H_{3}(x^2)+\frac{1}{2} \pi ^2
   H_{5}(x^2)+8 H_{7}(x^2) \nonumber \\
   &+\frac{1}{2} \pi ^2
   H_{2,3}(x^2)+9 H_{2,5}(x^2)+\pi ^2
   H_{3,2}(x^2)+14 H_{3,4}(x^2)+\frac{3}{2} \pi ^2
   H_{4,1}(x^2) \nonumber \\
   &+19 H_{4,3}(x^2)+25 H_{5,2}(x^2)+30 H_{6,1}(x^2)+6 H_{2,2,3}(x^2)+9 H_{2,3,2}(x^2)+9
   H_{2,4,1}(x^2) \nonumber \\
   &+8 H_{3,1,3}(x^2)+14 H_{3,2,2}(x^2)+14 H_{3,3,1}(x^2)+15
   H_{4,1,2}(x^2)+15 H_{4,2,1}(x^2) \nonumber \\
   &+12 H_{5,1,1}(x^2)-\frac{9 \zeta_7}{4}-\frac{9}{2}
   H_{2}(x^2) \zeta_5-\frac{\pi ^2 \zeta_5}{4}-\frac{1}{2} \pi ^2
   H_{2}(x^2) \zeta_3-4 H_{4}(x^2) \zeta_3 \nonumber \\
   &-6 H_{2,2}(x^2) \zeta_3-8 H_{3,1}(x^2) \zeta_3-\frac{\pi^4}{30}\zeta_3
\end{align}
}
at four loops. In region III, i.e.\ $x \in [-1,0]$, the logarithms are the only source of imaginary parts.
Together with the $i0$-prescription from section~\ref{sec-kinematics} the imaginary
part can therefore be extracted explicitly in this region.

\bibliographystyle{JHEP}

\providecommand{\href}[2]{#2}\begingroup\raggedright\endgroup

\end{document}